\documentclass[a4paper,11pt]{article}
\pdfoutput=1 

\usepackage{jcappub} 

\usepackage[T1]{fontenc} 
\usepackage[dvipsnames]{xcolor}
\usepackage{tikz}

\usepackage{amssymb, amsmath, epsfig, natbib}

\renewcommand{\vec}[1]{\mathbf{#1}}

\author[a,b]{Chirag Modi}
\author[a,b]{Martin White}
\author[c,d]{Zvonimir Vlah}

\affiliation[a]{Department of Physics, University of California,
Berkeley, CA 94720}
\affiliation[b]{Department of Astronomy, University of California,
Berkeley, CA 94720}
\affiliation[c]{Stanford Institute for Theoretical Physics and
Department of Physics, Stanford University, Stanford, CA 94306, USA} 
\affiliation[d]{Kavli Institute for Particle Astrophysics and Cosmology,
SLAC and Stanford University, Menlo Park, CA 94025, USA}

\emailAdd{modichirag@berkeley.edu}
\emailAdd{mwhite@berkeley.edu}
\emailAdd{zvlah@stanford.edu}

\title{Modeling CMB Lensing Cross Correlations with {\sc CLEFT}}


\abstract{A new generation of surveys will soon map large fractions of sky
to ever greater depths and their science goals can be enhanced by exploiting
cross correlations between them.  In this paper we study cross correlations
between the lensing of the CMB and biased tracers of large-scale structure at
high $z$.  We motivate the need for more sophisticated bias models for
modeling increasingly biased tracers at these redshifts and propose the use
of perturbation theories, specifically Convolution Lagrangian Effective
Field Theory ({\sc CLEFT}).  Since such signals reside at large scales and
redshifts, they can be well described by perturbative approaches.
We compare our model with the current approach of using scale independent bias
coupled with fitting functions for non-linear matter power spectra, showing
that the latter will not be sufficient for upcoming surveys.  We illustrate
our ideas by estimating $\sigma_8$ from the auto- and cross-spectra of mock
surveys, finding that {\sc CLEFT} returns accurate and unbiased results at
high $z$.  We discuss uncertainties due to the redshift distribution of the
tracers, and several avenues for future development.}
\arxivnumber{1706.NNNNN}

\begin{document}
\maketitle
\flushbottom

\section{Introduction}

In the last decade, gravitational lensing of the cosmic microwave background
(CMB) has arisen as a promising new probe of cosmology
(see Refs.~\cite{LewCha06,Han10} for reviews).
CMB photons are deflected by the gravitational potentials associated with 
large-scale structure (LSS) between us and the last scattering surface,
providing a probe of late-time physics directly in the CMB sky.
This effect is sensitive to the geometry of the universe and the growth and
structure of the matter distribution, making it a powerful probe of dark
energy, modifications to General Relativity and the sum of neutrino masses.
Relying on the well-understood statistics of the CMB anisotropies, with a
well defined and constrained source redshift, CMB lensing is immune to many
of the systematics that need to be modeled for cosmic shear surveys using
galaxies and is particularly powerful at $z\simeq 1-5$ where galaxy lensing
surveys become increasingly difficult.
This lensing effect has been robustly detected by multiple CMB experiments
\cite{Smi07,Hir08,vanE12,Ade14,vanE15,PlanckLens13,PlanckLens15},
with the most recent detections by Planck reaching $40\,\sigma$
and providing nearly full sky maps of the (projected) matter density all the
way back to the surface of last scattering.
In future, even more powerful experiments such as Advanced ACT \cite{deB16}
the Simons Observatory \cite{Suz16} and a Stage IV, ground based CMB experiment
(CMB S4; \cite{CMBS4}) will map larger fractions of the sky with greater
fidelity. 

As a community we are also investing in large scale imaging surveys such as the
Dark Energy Survey (DES\footnote{https://www.darkenergysurvey.org/}),
DECam Legacy Survey (DECaLS\footnote{http://legacysurvey.org}), 
Subaru Hyper Suprime-Cam (HSC\footnote{http://hsc.mtk.nao.ac.jp/ssp/}),
Large Synoptic Survey Telescope (LSST\footnote{https://www.lsst.org}),
Euclid\footnote{http://sci.esa.int/euclid}
and WFIRST\footnote{https://wfirst.gsfc.nasa.gov}
to map the sky to greater depths in multiple bands.
Imaging surveys which cover the same region of the sky as CMB surveys can
enhance their science return through joint analysis, for example by
cross-correlating the density field traced by one survey with that of
another.
Ideally such a cross-correlation can benefit from the strengths of the two
probes while being insensitive to the systematics that could plague either.

The study of cross-correlations of CMB lensing with other tracers of large
scale structures, such as galaxy surveys, enables tests of General Relativity,
probes the galaxy-halo connection, allows isolation of the lensing signal
in narrow redshift intervals and can give a handle on various systematics such
as biases in photometric redshifts \cite{Bax16}
or multiplicative biases in shear measurements
\cite{Val12,Val13,Das13,Liu16,Bax16,Sin17,Sch17}.
CMB lensing maps have been cross-correlated with galaxies and quasars
\cite{Smi07,Hir08,Ble12,She12,PlanckLens13,Geach13,OmoriHolder15,Ferraro15,
DiPompeo15,DiPompeo16,PlanckLens15,All15,Bianchini15,Pullen16,Giannantonio16}.
They have been cross-correlated with galaxy-based cosmic shear maps
\cite{LiuHill15,Miyatake16,Singh17,Bax16,Har16,Har17},
with the Ly$\alpha$ forest \cite{Doux16}
and with unresolved sources including dusty star-forming galaxies
\cite{PlanckLens13,PlanckLens15,Omori17}
and the $\gamma$-ray sky from Fermi-LAT \citep{Fornengo15,Feng17}.

As statistical errors from surveys decrease the level of sophistication of
the analysis and the accuracy of the models must increase. In particular,
in order to interpret CMB lensing-galaxy cross-correlation observations
we need a flexible yet accurate model for the clustering of both biased
tracers and the matter.
To date most analyses have used fitting functions for the non-linear, matter
power spectrum and a scale-independent linear bias.
These are reasonable approximations at the current level of precision, however
as the statistical errors decrease the model must be improved. 
Since the CMB lensing is most sensitive to structure at high redshifts
($z\simeq 1-5$), and at relatively large scales, higher order perturbation
theory seems a natural choice for this modeling. 
The perturbative approach, and the need for sophisticated bias modeling, will
only become more relevant as imaging surveys probe ever higher redshifts and
ever more sources.

The focus of this paper will thus be on modeling the cross-correlation of CMB 
lensing with biased tracers (halos), and their auto-correlations  using 
perturbation theory.  In particular we use Lagrangian perturbation theory and 
effective field theory, coupled with a flexible Lagrangian bias model,
which makes accurate predictions for large-scale auto- and cross-correlations
in both configuration and Fourier space
(see e.g.~Ref.~\cite{VlaCasWhi16}, building upon the work of
 Refs.~\cite{Buc89,Mou91,Hiv95,Mat08a,Mat08b,CLPT,PorSenZal14,McQWhi15,
 VlaSelBal15,VlaWhiAvi15}). 
The outline of the paper is as follows.  In  \S\ref{sec:background} we review
some background material on CMB lensing as well as our perturbation theory
model and establish our notation.
We also discuss the instrumental noise and sampling variance in 
future surveys  which sets the error budget for our modeling. 
In \S\ref{sec:nbody} we use N-body simulations to gauge the performance of
our model.  In \S\ref{sec:Pmm} we give an example of how CMB lensing
cross-correlations can constrain cosmological parameters by estimating the
power spectrum amplitude, $\sigma_8$, from our N-body data.
We compare our model against the current approach of using a fitting function
for the non-linear, matter power spectrum with a scale dependent bias.
We look at how measurement errors and parameter marginalization affect this
measurement in \S\ref{sec:dndz}.
Finally, we conclude with a discussion in \S\ref{sec:conclusions}.
We discuss a simplified perturbative model, appropriate for near-future data
analysis, and our forecasting methodology in the appendices.
Throughout we shall use comoving coordinates and assume spatially
flat hypersurfaces.  Where we need to assume a cosmology we use the same
cosmology as our N-body simulations (described in \S\ref{sec:nbody}).

\section{Background}
\label{sec:background}

\subsection{The angular power spectrum}
\label{sec:cl}

The photons which we see as the cosmic microwave background must traverse
the gravitational potentials associated with large scale structure between
us and the surface of last scattering.  These potentials cause the photons'
paths to be deflected, an effect known as gravitational lensing
\cite{LewCha06,Han10}.
Lensing remaps the temperature and polarization fields at $\hat{n}$ by an
angle $\vec{\alpha}=\nabla\psi$ where $\psi$ is the lensing potential (we
shall make the Born approximation throughout, so the $\psi$ is a weighted
integral of the Weyl potential along the line of sight).
We shall work in terms of the lensing convergence, $\kappa$, which is related
to $\psi$ through
$\kappa(\hat{n})=(-1/2)\nabla^2\psi(\hat{n})$ or
$\kappa_\ell=(1/2)\ell(\ell+1)\psi_\ell$.
We shall comment upon these approximations further below.

Both $\kappa$ and our tracer density are projections of 3D density fields.
We define the projection through kernels, $W(\chi)$, with $\chi$ the
line-of-sight distance.
Given two such fields on the sky the multipole expansion of the angular
cross-power spectrum is
\begin{equation}
  C_\ell^{XY} = \frac{2}{\pi}\int_0^\infty
  d\chi_1\,d\chi_2\ W^X(\chi_1)W^Y(\chi_2) \int_0^\infty k^2\,dk
  \ P_{XY}(k;z_1,z_2)j_\ell(k\chi_1)j_\ell(k\chi_2)
  \quad .
\end{equation}
Our focus will be on small angular scales (high $\ell$), where the signal
to noise is highest and the effects of quasi-linear evolution become important.
This allows us to make the Limber approximation, which in our context is
\begin{equation}
  \int k^2\,dk\ j_\ell(k\chi_1)j_\ell(k\chi_2) \approx \frac{\pi}{2\chi_1^2}
  \delta(\chi_1-\chi_2) \qquad .
\end{equation}
In this limit $C_\ell$ reduces to a single integral along $\chi$ of the
equal-time, real-space power spectrum:
\begin{equation}
  C_\ell^{XY} = \int d\chi\ \frac{W^X(\chi)W^Y(\chi)}{\chi^2}
  \ P_{XY}\left(K=\frac{\ell+1/2}{\chi},k_z=0\right)
\label{eqn:ClXY}
\end{equation}
where we have included the lowest order correction to the Limber approximation,
$\ell\to\ell+1/2$, to increase the accuracy to $\mathcal{O}(\ell^{-2})$
\cite{LovAfs08,ShaLew08}.  For the case of interest
\begin{equation}
  W^{\kappa}(\chi) = \frac{3}{2}\Omega_mH_0^2(1+z)
  \ \frac{\chi(\chi_\star-\chi)}{\chi_\star}
  \quad , \quad
  W^{g}(\chi) \propto H(z)\,\frac{dN}{dz}
\end{equation}
with $\chi_\star$ the (comoving) distance to last scattering and
$\int W^{g}d\chi=1$.  For ease of presentation we have neglected a
possible contribution from lensing magnification, which could be included
in $W^g$ if necessary.  Including this term does not materially affect our
later discussion or results.

For the convergence auto-power spectrum the integral extends to low $\chi$
and thus high $k$ where linear theory is no longer adequate and perturbation
theories are not quantitatively reliable \cite{ForSen16} (but see
\S\ref{sec:Pmm} for further discussion).  However, if we cross-correlate the
lensing signal with a tracer (e.g.~galaxy or quasar) which is localized at
high $z$ the low-$\chi$ cut-off in $W^g$ will reduce the sensitivity of
$C_\ell^{\kappa g}$ to high-$k$ physics.
In combination with the reduction in non-linear evolution at high $z$ this
motivates our use of perturbation theory for $P_{\kappa g}$.

In the limit that the tracer sample is well localized in redshift the angular
power spectrum is just proportional to the cross-power spectrum evaluated at
$\ell+1/2=k\chi_g$:
\begin{equation}
  C_\ell^{\kappa g} \approx \frac{W^\kappa(\chi_g)}{\chi_g^2}
  \ P_{\kappa g}\left(k=\frac{\ell+1/2}{\chi_g}\right) =
  \frac{3}{2}\Omega_mH_0^2(1+z)
  \,\frac{(\chi_\star-\chi_g)}{\chi_\star\chi_g}
  \ P_{\kappa g}\left(k=\frac{\ell+1/2}{\chi_g}\right)
  \quad .
\label{eqn:thin_slice}
\end{equation}
For tracers at distances of a few $h^{-1}$Gpc, e.g.~$z>1$,
even $\ell\sim 10^3$ corresponds to $k<1\,h\,{\rm Mpc}^{-1}$ which is
within the reach of perturbation theory at high $z$.
Similarly $C_\ell^{gg}\simeq P(k=[\ell+1/2]/\chi)/[\chi^2\Delta\chi]$
for a top-hat bin of width $k^{-1}\ll\Delta\chi\ll\chi$.

\begin{figure}
\begin{center}
\resizebox{\columnwidth}{!}{\includegraphics{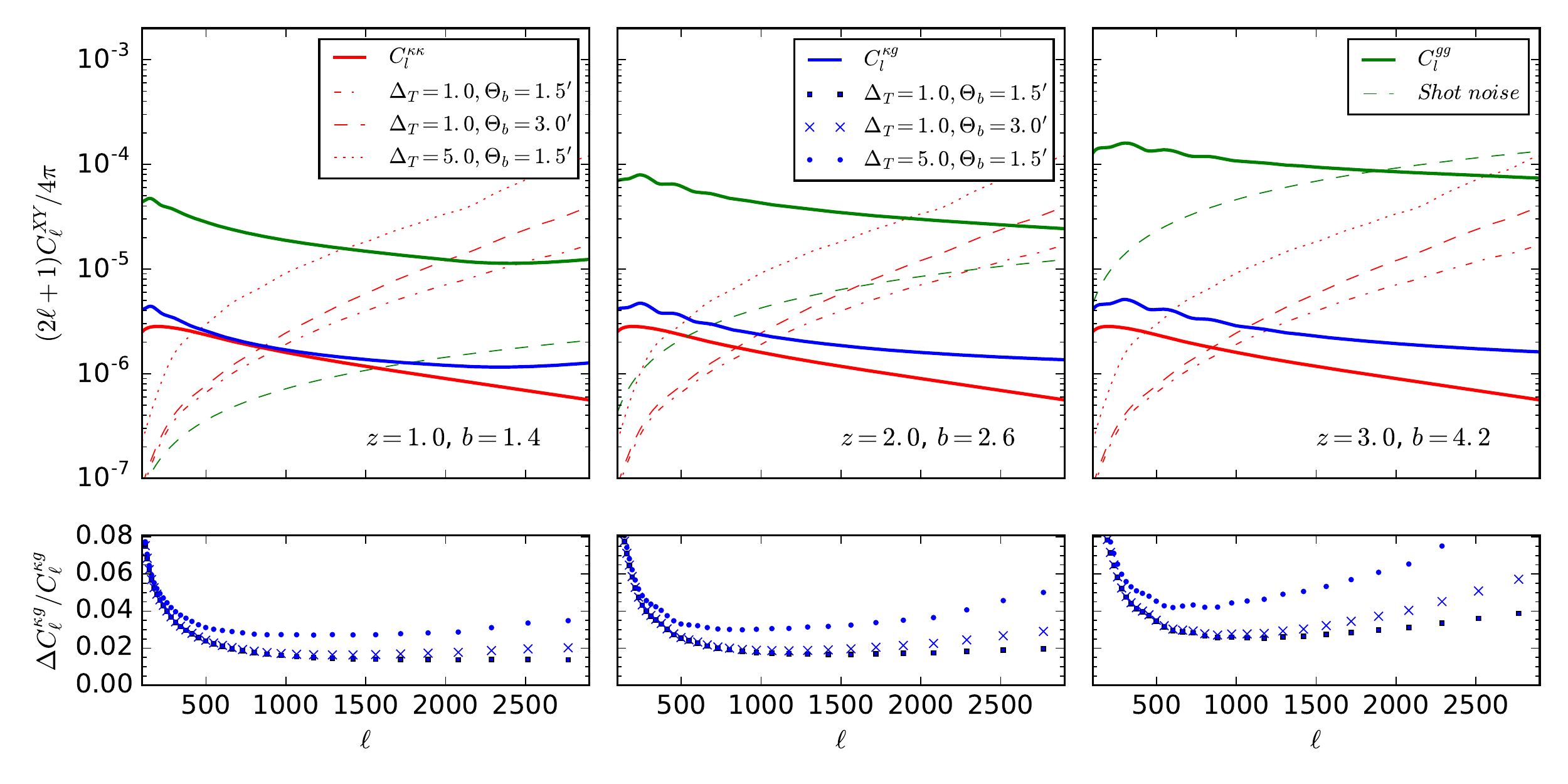}}
\end{center}
\caption{The signal and noise angular power spectra at $z=1$ (left), $z=2$
(middle) and $z=3$ (right).  \textit{Upper panels}: the power spectra for
the lensing and galaxy auto-correlations ($\kappa\kappa$ and $gg$) and
the cross-correlation ($\kappa g$) for a bin of width $\Delta z=0.5$ and
different combinations of noise and beam sizes.
The galaxy auto-correlations ($C_\ell^{gg}$) assume the halo power spectra
of our N-body simulation (\S\ref{sec:nbody}) but a shot-noise appropriate to
the LSST gold sample.
The lensing noise is the minimum variance combination of $TT$, $TE$, $EE$
and $EB$ as described in Appendix \ref{app:noise}.
We have assumed the CMB and galaxy survey overlap on 50\% of the sky.
\textit{Lower panels:} The fractional error on $C_\ell^{\kappa g}$ for bins
of $0.1\,\ell$.  Future experiments could approach 1\% precision on
$C_\ell^{\kappa g}$ in multiple bins.}
\label{fig:cl}
\end{figure}

Fig.~\ref{fig:cl} shows the signal and noise angular power spectra as
well as the inferred fractional error on the cross-correlation,
$C_\ell^{\kappa g}$, for some example configurations.
We have used the CLASS code \cite{CLASS} to compute the CMB lensing spectra.
To make contact with later sections, we have taken the tracer signal levels
appropriate for halos of $10^{12}\,h^{-1}M_\odot$ in our N-body simulation
(see \S \ref{sec:nbody}) but we use the $dN/dz$ of the LSST gold sample
\cite{LSST} in slices of width $\Delta z=0.5$.

The assumptions and formalism used to estimate the uncertainties is described
in Appendix \ref{app:noise}.  In particular the lensing noise is the minimum
variance combination of $TT$, $TE$, $EE$ and $EB$, with $\ell_{\rm max}=3000$,
5000 for temperature and polarization respectively.
We find that, for a given noise level, the errors in $C_\ell^{\kappa\kappa}$
and $C_\ell^{\kappa g}$ are quite insensitive to angular resolution in the
range $1'-3'$ FWHM (see also Ref.~\cite{Sch17}).
The $TT$ contribution also stays fairly constant with
map noise levels between $1-5\,\mu$K-arcmin (after foreground cleaning).
Below approximately $2\,\mu$K-arcmin noise the $EB$ contribution begins to
significantly reduce the uncertainty on $\kappa$.
Next-generation CMB experiments are noise-limited in lensing, per $\ell$,
beyond $\ell$ of a few hundred but there is still significant constraining
power at high $\ell$ because of the many modes which can be averaged together.
An experiment such as CMB-S4 would be sample variance limited to just below
$\ell=10^3$.

The uncertainty in the cross correlation has contributions from CMB map noise
and shot noise in the imaging survey.  As for the $\kappa$ noise, this is also
fairly inensitive to the beam for scales larger than $\ell = 2000$,
but the fractional errors increase by more than $1.5\%$ for $\ell>1500$
on increasing the map noise from $1\,\mu$K-arcmin to $5\,\mu$K-arcmin.
For the cases shown in Fig.~\ref{fig:cl} the shot-noise is highly subdominant
at lower $z$ and so the fractional cross correlation uncertainty is very
weakly dependent on the level of shot noise (it would increase by only
$\sim 0.5\%$ for a survey one magnitude shallower).
However the errors start to depend upon shot noise at the higher redshifts.
We note that despite averaging modes in bins of $\Delta\ell=0.1$, the
fractional error in $C_\ell^{\kappa g}$ reaches a minimum of $1\%$ around
$\ell = 1500$ at $z=2$ and then starts to increase again.
This thus sets the minimum level of accuracy that we need from our model.
 
Current generation large scale surveys, such as DES, are completely dominated
by shot noise at $z=2$ and $z=3$ on scales smaller than $\ell=1000$.
Deeper surveys, such as HSC, suffer primarily due to smaller sky coverage
and increased sample variance.
A future survey like LSST has the combination of depth and area to provide
strong constraints to $\ell=1000$ at $z=1$ and $2$, with shot noise becoming
important only at higher $z$.
There is still significant CMB lensing contribution at high redshift, however,
and thus significant potential constraining power.
It is therefore worthwhile considering alternative techniques to improve SNR
when pushing to higher redshifts.
At these redshifts one can efficiently select samples of galaxies using
dropout techniques, e.g.~$u$-band dropouts for $z\sim 3$ and $g$-band
dropouts for $z\sim 4$.  Magnitude limited dropout samples naturally
produce bands in redshift of about $\Delta z\sim 0.5$ with clustering
properties that are similar to normal galaxies at $z=0$ \cite{Ste03,Ade05}.
Using the UV luminosity functions of Ref.~\cite{Red08} reaching a number
density at which we are sample variance limited (in $C_\ell^{gg}$) to
$\ell=2000$ for our $10^{12}\,h^{-1}M_\odot$ halos requires an
$\mathcal{R}$-band depth of about $24.3$.
It might be more efficient to look at $g$-band dropouts (i.e.~$z\sim 4$),
where it is possible to go fairly deep relatively quickly in the dropout
band.
Another alternative would be medium or narrow band surveys which are
targeted at specific redshift ranges, picking up e.g.~Lyman-$\alpha$
emitting galaxies.
It is not the purpose of this paper to propose a deeper imaging survey,
so we leave this topic for future investigation.
Rather we shall take the above to suggest it is possible to achieve
percent-level constraints on the cross-correlation over a wide range of
$\ell$ and $z$ from future experiments (or by enhancing current surveys).
This motivates our development of an appropriate theory for the interpretation
of such data.

Throughout this paper we shall follow standard practice and approximate
the lensing using the Born approximation, though we shall include
non-linear terms in the large-scale densities.
As the precision improves it will be necessary to reconsider all such
approximations \cite{Bohm16,LewPra16} for cross-correlations as well as
the auto-spectrum of $\kappa$ and to worry about cleaning out contaminants
\cite{Amb04,vanE12,FerHil17}.
Isolating the signal to higher redshift, where the non-linearity is less
pronounced, makes the cross-spectrum less sensitive to bispectrum and
trispectrum terms than the $\kappa$ auto-spectrum.
However by focusing on overdense regions where biased tracers reside the
impact of non-linearities is enhanced.
How this impacts a cross-correlation measurement from a lensed CMB sky
will require further investigation.

\subsection{Lightcone evolution: the effective redshift}
\label{sec:zeff}

Once the evolution of $P(k,z)$ is specified, the theory of \S\ref{sec:cl}
can be used to provide an accurate prediction for the auto- and cross-
angular power spectra which are observed.
This allows us to compare theory and observation even for sources with
broad redshift kernels where we expect significant evolution across the
sample.  However it is often the case that we wish to interpret the $C_\ell$,
which involve integrals across cosmic time, as measurements of the clustering
strength at a single, ``effective'', epoch or redshift.
Motivated by Eq.~(\ref{eqn:ClXY}) we define
\begin{equation}
  z_{\rm eff}^{XY}=\frac{\int d\chi\ \left[W^X(\chi)W^Y(\chi)/\chi^2\right]\ z}
  {\int d\chi\ \left[W^X(\chi)W^Y(\chi)/\chi^2\right]\ \hphantom{z}}
\end{equation}
such that the linear term in the expansion of $P(k,z)$ about
$z_{\rm eff}^{XY}$ cancels in the computation of $C_\ell^{XY}$.

We have compared $C_\ell^{\kappa g}$ and $C_\ell^{gg}$ computed using an
evolving $P(k,z)$ to that produced by using $P(k,z_{\rm eff})$ in
Eq.~(\ref{eqn:ClXY}) for several $dN/dz$ shapes and widths, $\Delta z$.
Using the evolution of the halo sample of \S\ref{sec:nbody} as an example
we find the $C_\ell$ are within 1.5\% for $\Delta z\le 0.5$ and $\ell>10$
for $1<z<3$.  The difference rises quickly beyond $\Delta z=0.5$ and is 5\%
for $\Delta z=1$.
The evolution of $P_{hh}$ we have used as an example is quite strong, since
we have focused on a fixed halo mass and thus a tracer whose bias increases
strongly with redshift.  More gradual evolution of $P(k)$ (e.g.~passive
evolution) would obviously lead to smaller effects, with no effect in the
limit of constant $P(k)$.

In what follows we shall use $\Delta z=0.5$ and approximate our inferences
as $P(k,z_{\rm eff})$.  Obviously as the width of the slice is reduced the
angular clustering of the tracers is enhanced and the approximation of
$P(k,z)$ by $P(k,z_{\rm eff})$ improves.  However in this limit the shot
noise increases as well, and the correlation with the $\kappa$ field
decreases rapidly.  For increases in $\Delta z$, the cross-correlation
increases and the shot-noise drops (as does the signal) but we trade 
the possibility of multiple independent thin slices in redshifts which can
be combined to reduce errors (in quadrature), to a single thick slice with
smaller errors.  We find that error in the auto-spectra are smaller when
using a single thick slice while those in cross spectrum prefer multiple
thin slices slices.  This opposing trend, coupled with the caveat that one
needs a model for the evolution of $P(k)$ in order to interpret the
observations from thick slices, makes $\Delta z =0.5$ a suitable choice
for our current work.

\subsection{Convolution Lagrangian effective field theory (CLEFT)}
\label{sec:lpt}

As argued above, we desire a flexible yet accurate model for the auto- and
cross-clustering of biased tracers and the matter in order to exploit the
information soon to be available from surveys.
Since the observations probe high redshift and relatively large scales,
higher order perturbation theory seems a natural choice.
In particular Lagrangian perturbation theory and effective field theory,
coupled with a flexbile Lagrangian bias model, offer a systematic and
accurate means of predicting the clustering of biased tracers in both
configuration and Fourier space (e.g.~Ref.~\cite{VlaCasWhi16}) making it
an ideal tool for modeling cross-correlations of CMB lensing with tracers
of large-scale structure.  Below we shall present only the Fourier space
formalism for brevity, though in some instances configuration space analyses
may be preferred.  Our formalism naturally handles both views with the
same parameters \cite{VlaCasWhi16} so it can be employed in fitting data in
either space.

\begin{figure}
\begin{center}
\resizebox{\columnwidth}{!}{\includegraphics{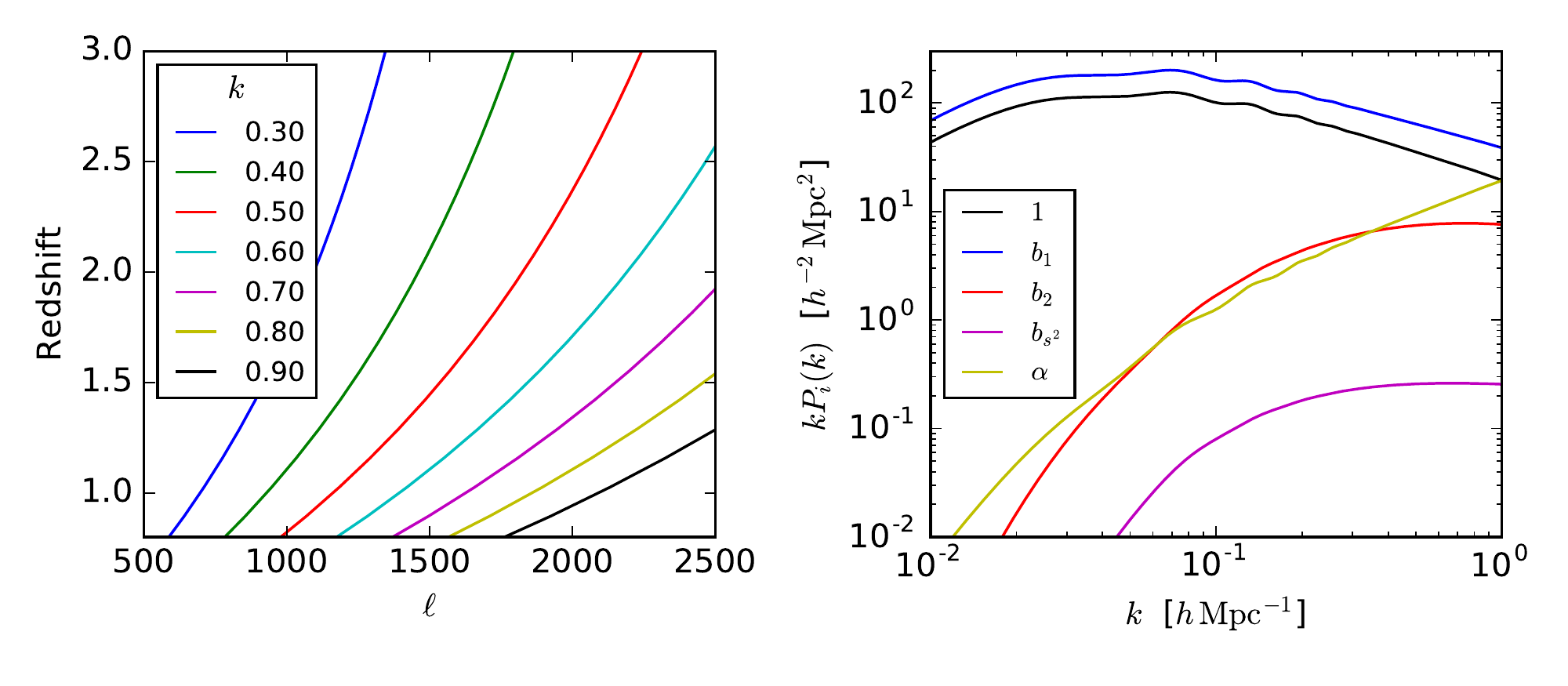}}
\end{center}
\caption{(Left) The mapping between $k$ [in $h\,{\rm Mpc}^{-1}$] and $\ell$
as a function of redshift for the cosmology of our N-body simulation.  In
the range $1<z<3$, which is our focus, angular scales $\ell<10^3$ correspond
to the quasi-linear scales easily within the reach of perturbation theory.
(Right) Contributions of the various terms in Eq.~(\ref{eqn:pcross}) at $z=2$
using the best-fit parameters determined in \S\ref{sec:nbody}.}
\label{fig:kstuff}
\end{figure}

The cross-correlation between the matter and a biased tracer, in real
space, contains a subset of the terms described in Ref.~\cite{VlaCasWhi16}.
Specifically the cross-power spectrum can be expressed as\footnote{In this
paper we shall not consider the effects of massive neutrinos, but for small
neutrino masses they can be easily included in our formalism by using only
the cold dark matter plus baryon linear power spectrum when computing the
CLEFT predictions and then adding in the linear neutrino power spectrum with
mass weighting.}
\begin{equation}
  P_{mg}(k) =
         \left(1-\frac{\alpha_\times\,k^2}{2}\right)P_Z + P_{\rm 1-loop}
       + \frac{b_1}{2}P_{b_1}+\frac{b_2}{2}P_{b_2}
       + \frac{b_{s^2}}{2}P_{b_{s^2}}
       + b_{\nabla^2}P_{b_{\nabla^2}}
       + s_\times
\label{eqn:pcross}
\end{equation}
where $P_Z$ and $P_{\rm 1-loop}$ are the Zeldovich and 1-loop matter terms,
the $b_i$ are Lagrangian bias parameters for the biased tracer,
$\alpha_\times$ is a free parameter which accounts for small-scale physics
not modeled by LPT and $s_\times$ is a possible ``stochastic'' contribution.
The individual $P_x$ can be written as spherical Hankel transforms
\begin{equation}
  P_x = 4\pi\int q^2\,dq\ e^{-(1/2)k^2(X_L+Y_L)}
  \left[f_x^{(0)}(k,q)j_0(kq) + \sum_{n=1}^\infty f_x^{(n)}(k,q)
  \left(\frac{kY_L}{q}\right)^nj_n(kq)\right]
\label{eqn:Px}
\end{equation}
with the linear Lagrangian correlator decomposed as
$A_{ij}^{\rm lin}=\delta_{ij}X_L+\hat{q}_i\hat{q}_jY_L$
and the $f_x^{(n)}$ given in \cite{VlaWhiAvi15,VlaCasWhi16}
(see Appendix \ref{app:hz} for more details and a simplified model).
All of these results assume that the LPT kernels are time-independent.
This is an excellent approximation for the density fields at high redshift
that we consider \cite{Ber02,Tak08,FasVla16}.

For the halo auto-spectrum the stochastic term includes a contribution from
shot noise and can be taken to be scale-independent at the order we work
(i.e.~a constant).  We find that this term is very well predicted by a Poisson
shot noise and since we subtract such a term from our ``signal'' spectra
(\S\ref{sec:nbody}) we can omit it.
For the matter-halo cross-spectrum the stochastic term scales as $k^2$ as
$k\to 0$ (but is unconstrained at high $k$) and is also generally omitted.
We have experimented with different forms and values of $s_\times$
and find our results are not particularly sensitive to such choices.  The fit
is slightly improved if we include a constant or a form like
$(k/k_\star)^2/[1+(k/k_\star)^2]$ with
$k_\star\simeq 0.1-0.5\,h\,{\rm Mpc}^{-1}$.
This amplitude of this term is never particularly large, and it helps
primarily at high $k$.
We choose to also omit this term for simplicity, though we note that including
an additional constant as a nuisance parameter could help when fitting data.
It is also worth noting that in the N-body simulations to which we compare
in \S\ref{sec:nbody} we may have an additional contribution from the finite
sampling of the density field by dark matter particles.
A Poisson contribution to the cross-spectrum,
$(\bar{n}_{\rm halo}\bar{n}_{\rm dm})^{-1/2}$,
would be in the range $10-30\,h^{-3}\,{\rm Mpc}^3$ for the samples we discuss
in \S\ref{sec:nbody} and thus not negligible at high $k$.  Thus when comparing
to the N-body we could have an additional contribution to $P_{mg}$ which
is constant (to lowest order) and potentially as large as the Poisson value
above.  We assume henceforth that this term is negligble.  Clearly a better
understanding of the stochastic terms could yield benefits in pushing the
model to higher $k$, but awaits further theoretical developments.

The $b_i$ represent bias terms
(for a recent review of bias see Ref.~\cite{Des16}, for a discussion of the
 advantages of a Lagrangian approach see Ref.~\cite{VlaCasWhi16}).
The lowest order term, $b_1$, dominates on large scales and is related to
the ``linear'', scale-independent, Eulerian bias $b=1+b_1$.
The second term, $b_2$, encodes scale-dependence while $b_{s^2}$ and
$b_{\nabla^2}$ encode the dependence of the object density on second
derivatives of the linear density field (e.g.~a constraint that objects
form at peaks).  We find we do not need these last two terms at high
redshift where our theory performs best, but they could become important
to accurately model clustering at higher $k$ \cite{Des16}.
These additional terms could also become more important for samples where
assembly bias plays a role, or samples with specific kinds of formation
histories, e.g.~galaxies selected via color cuts, with strong emission lines
or which have more reliable photometric redshifts.

Within the peak-packground split for the Press-Schechter mass function
\cite{PreSch74} the first two bias parameters are related to the peak
height, $\nu$, and the critical density for collapse, $\delta_c$, by
\begin{equation}
  b_1 = \frac{\nu^2-1}{\delta_c} \qquad , \qquad
  b_2 = \frac{\nu^4-3\nu^2}{\delta_c^2}
  \label{eq:ps_bias}
\end{equation}
In this model $b=1$ would correspond to $b_1=0$ and $b_2=-0.7$, $b=2$ to
$b_1=1$ and $b_2=-0.3$.
Note that $b_2\to b_1^2$ as $b_1\to\infty$, so the scale-dependence of the bias
is predicted to become more pronounced as the bias increases.
This expectation is borne out in our fits, however we find that the
relationship between $b_1$ and $b_2$ is not quantitatively very accurate so
we treat the $b_i$ as free parameters.
There is some evidence from N-body simulations that a relationship between
the $b_i$ does exist \cite{Laz16,Cas17,Mod17}.
Using such relationships as priors on the parameters could yield benefits
for some science goals, as we discuss later.
The derivative bias terms, $b_{s^2}$ and $b_{\nabla^2}$, only become important
on small scales and we shall not include them below.

The expression for the auto-spectrum of the biased tracers can be found in
Ref.~\cite{VlaCasWhi16} and we shall not reproduce it here.  In addition to
the terms linear in $b_1$, $b_2$, etc.~it contains quadratic terms like
$b_1^2$, $b_1b_{s^2}$, and so on.  The bias terms, $b_i$, are common to the
auto- and cross-spectra but the value\footnote{The $\alpha$ coefficient
represents a degenerate combination of the effects of small-scale physics
and scale-dependent bias.} of $\alpha$ can be different for each spectrum.
We denote these as $\alpha_\times$ and $\alpha_a$, with the subscript $a$
referring to the auto-spectrum.

As we explore below, when comparing to observations involving biased tracers
choosing a sophisticated and flexible bias model is essential in order not to
introduce errors.
In fact the impact of beyond-linear bias parameters is equal in importance to
the effects of non-linear gravitational evolution.
In the language of EFT, the ``cut-off'' scale associated with biasing is of
order the Lagrangian radius of the halos hosting our tracers.  For a fixed
halo mass this is a redshift-independent scale.  By contrast the cut-off
associated with gravitational non-linearity moves to higher $k$ at higher $z$.

Fig.~\ref{fig:kstuff} shows the relative contribution to $P_{mh}$ of
different terms at $z=2$, using the best fit parameters determined in the
next section.
We can see that the dominant terms are $P_Z$, $P_{\rm 1-loop}$ and $P_{b_1}$.
The other terms are subdominant, but can affect the predictions at the high
accuracy which will be demanded by future observations.

\section{Comparison with N-body simulations}
\label{sec:nbody}

In order to validate our approach, we compare our analytic models to the
cross-power spectrum between halos and dark matter measured in N-body
simulations.  For this purpose we make use of 10 simulations run with the
TreePM code of \cite{TreePM}.  Each simulation employs the same ($\Lambda$CDM)
cosmology but with a different random number seed chosen for the initial
conditions.  These simulations have been described in more detail elsewhere
\cite{Rei14,RSDmock,VlaWhiAvi15}, but briefly they were performed in
boxes of size $1380\,h^{-1}$Mpc with $2048^3$ particles and modeled a
$\Lambda$CDM cosmology with $\Omega_m=0.292$, $h=0.69$, $n_s=0.965$
and $\sigma_8=0.82$.
We use outputs at $z=1$, $2$ and $3$ to sample the range of most interest
for cross-correlations with CMB lensing.
For each output we compute the real-space auto-spectra of the halos and
matter and the cross-spectrum between the halos and matter for
friends-of-friends ($b=0.168$) halos with
$10^{12.0}<M<10^{12.5}\,h^{-1}M_\odot$.
The power spectra were computed on a $2048^3$ grid, using cloud-in-cell
interpolation, and the spectra were corrected for the window function of
the charge assignment and for (Poisson) shot-noise.
The number density of halos was $1.70\times 10^{-3}\,h^3\,{\rm Mpc}^{-3}$ at
$z=1$, $1.0\times 10^{-3}\,h^3\,{\rm Mpc}^{-3}$ at $z=2$ and
$3.8\times 10^{-4}\,h^3\,{\rm Mpc}^{-3}$ at $z=3$.

\begin{figure}
\begin{center}
\resizebox{\columnwidth}{!}{\includegraphics{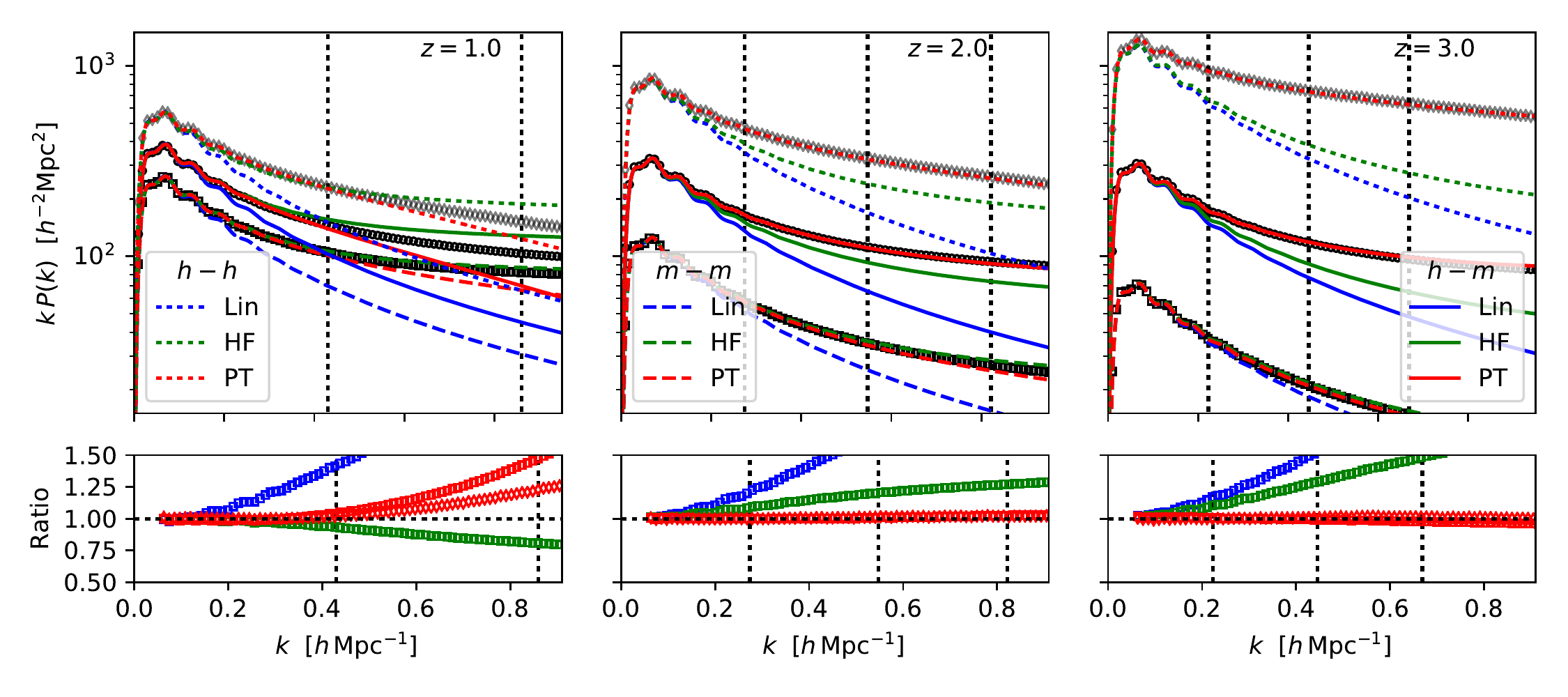}}
\end{center}
\caption{A comparison of the analytic model to the results from N-body
simulations.  The upper panels show $k\,P(k)$ for the matter auto-spectrum
(lower set of dashed lines and squares) the halo-matter cross spectrum
(middle set of solid lines and circles) and the halo-halo auto-spectrum
(upper set of dotted lines and diamonds) with shot-noise subtracted.
The points show the N-body results (in real space) at
(left) $z=1$, (middle) $z=2$ and (right) $z=3$.
Blue lines show the linear theory with a constant bias, the green lines show
the {\sc HaloFit} \cite{Tak12} spectra with constant bias while the red lines
show the perturbation theory (PT) predictions.
In the lower panel we show the ratio of the N-body cross-spectra ($P_{mh}$)
to each of the linear theory, {\sc HaloFit} and PT predictions on an expanded
$y$-scale.  For the PT predictions we also show the ratio for the auto-spectra
(red diamonds).  The vertical dotted lines mark $\ell=1000$, $2000$ and $3000$
[missing in the $z=1$ panel].}
\label{fig:nbody}
\end{figure}

Fig.~\ref{fig:nbody} compares the N-body results to the CLEFT results
of the previous section, and to the {\sc HaloFit}\footnote{We
use the implementation in CLASS.} fitting function \cite{Tak12}.
Upper panels show the comparison over the full range while the lower panels
show the ratio of the N-body results to each of the theoretical models, with
an expanded $y$-axis scale to highlight small deviations.
In the upper panel the squares show the matter power spectrum, where we see
that the N-body departs significantly from linear theory even at
$k\simeq 0.25\,h\,{\rm Mpc}^{-1}$: $P/P_L=1.15$, $1.07$ and $1.04$.
This is consistent with the level of power, as measured by the dimensionless
power spectrum: $\Delta^2(k=0.25\,h\,{\rm Mpc}^{-1},z)=0.45$, $0.20$ and $0.11$
at $z=1$, $2$ and $3$.  Another measure of the non-linear scale is the 1D, rms
Zeldovich displacement, $\Sigma$.
At $k=0.25\,h\,{\rm Mpc}^{-1}$ the product $k\Sigma$ is $0.9$, $0.6$ and $0.5$
at $z=1$, $2$ and $3$.
Unlike linear theory, the agreement between 1-loop perturbation theory and the
N-body results is very good to quite high $k$: within 1\% out to
$k=0.3$, $0.4$ and $0.6\,h\,{\rm Mpc}^{-1}$ for $z=1$, 2 and 3 and within 5\%
to $k=0.5\,h\,{\rm Mpc}^{-1}$ at $z=1$ and $k\simeq 0.7\,h\,{\rm Mpc}^{-1}$
at $z\ge2$.
For comparison the updated {\sc HaloFit} fitting function \cite{Tak12}
fits the N-body matter power spectrum almost within the quoted accuracy
(5\% for $k<1\,h\,{\rm Mpc}^{-1}$ and $0<z<10$) with a maximum deviation
of $6\%$ in the $z=3$ output.  A recent comparison of the performance of
different fitting functions for the matter power spectrum can be found in
Ref.~\cite{Law17}.

The results of direct relevance for our purposes are the halo-matter
cross-correlations and the halo-halo auto-correlations, also shown in
the upper panel of Fig.~\ref{fig:nbody}.
We perform a joint fit to the two spectra, so that the relevant bias terms
are self-consistent.  Unlike later sections, in these fits we put most of the
weight at low $k$ (enforcing a good match at low $k$ and to reduce over-fitting)
and we allow $b_{s^2}$ to be free to test it has a small impact.
Concentrating on the cross-spectrum, the lower panels show the ratio
N-body/model for CLEFT and the newer {\sc HaloFit} of Ref.~\cite{Tak12}
with an expanded $y$-axis.
We see that the best-fitting perturbation theory model matches the N-body
data at the few percent level out to $\ell\simeq 750$ for $z=1$ and to
$\ell\simeq 2000$ for $z=2$ and 3\footnote{There is likely some degree of
over-fitting in the cross-correlation results of Fig.~\ref{fig:nbody},
since we would not expect the cross-spectrum to fit well to higher $k$ than
the matter auto-spectrum.  Even so it seems that CLEFT provides a percent-level
accurate method for predicting the halo-matter cross-power spectrum to
$\ell=1000$ and possibly $\ell=2000$.}.
Key to this level of agreement is a flexible bias model.
The constant bias, linear theory results are not accurate for
$k\ge 0.1\,h\,{\rm Mpc}^{-1}$ even at these high redshifts.
{\sc HaloFit} improves over linear theory by quite a bit, but a
scale-independent bias is not a good model for the clustering of these
halos at $1<z<3$ even at $\ell<10^3$.
The errors introduced by assuming a scale-independent bias can exceed 10\%
on quasi-linear scales.

\begin{figure}
\begin{center}
\resizebox{3in}{!}{\includegraphics{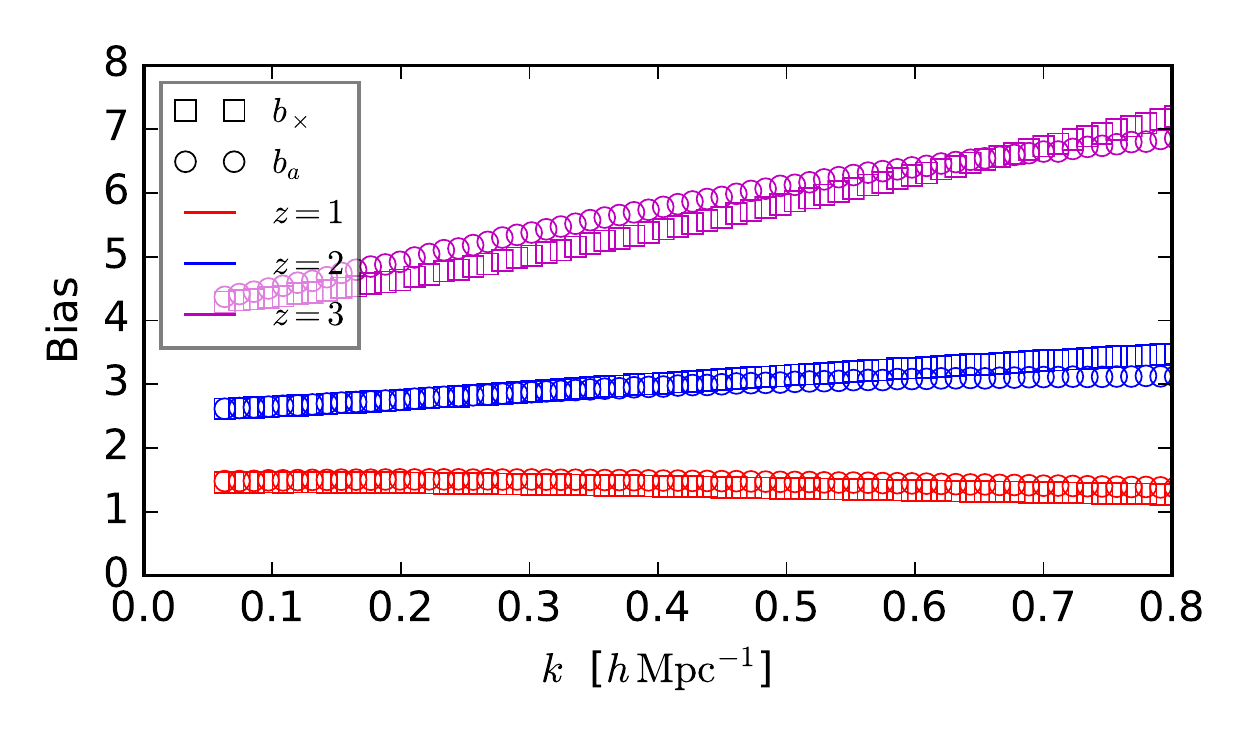}}
\resizebox{3in}{!}{\includegraphics{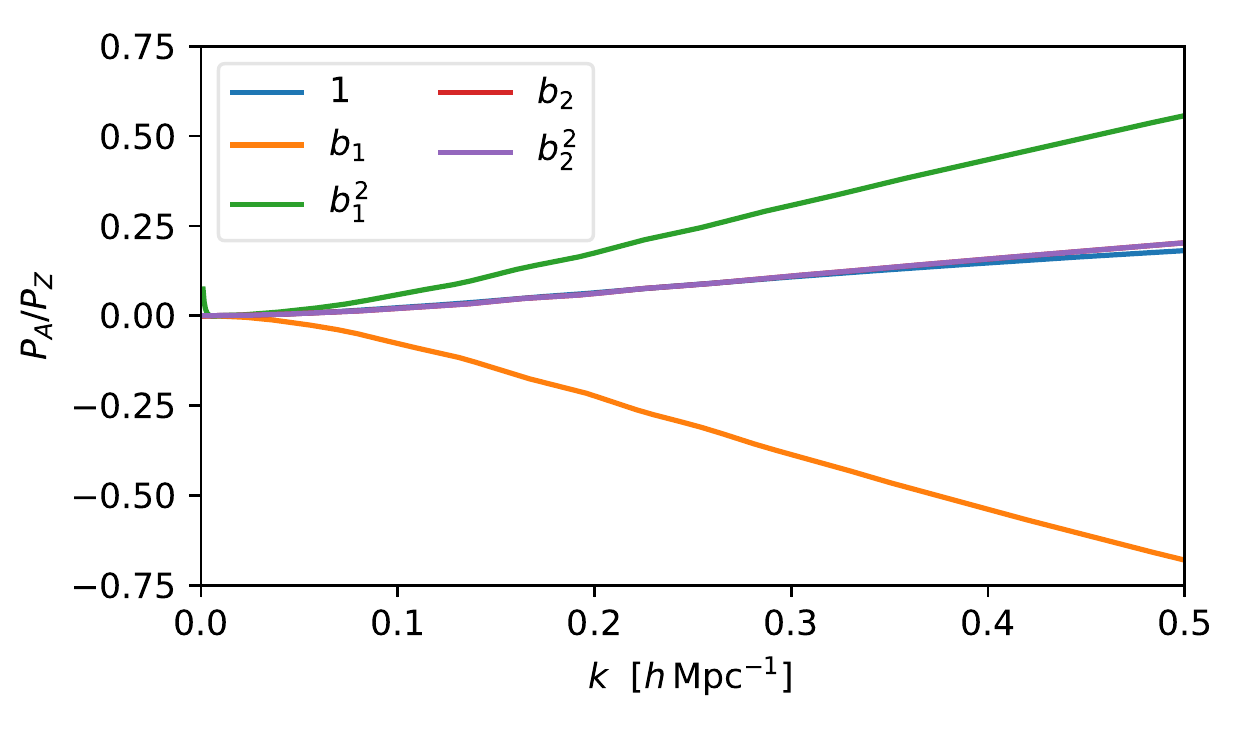}}
\end{center}
\caption{(Left) The bias terms, estimated from the cross- and auto-spectra,
for our N-body halos at $z=1$, $2$ and $3$.  Motivated by
Eq.~(\ref{eqn:halofit_peaks}) we define
$b_\times(k)=P_{hm}(k)/P_{mm}(k)$ and $b_a(k)=[P_{hh}(k)/P_{mm}(k)]^{1/2}$.
Note that the bias is scale dependent, but the scale dependence is
different for the cross- and auto-spectra.  Both the scale dependence and the
difference become more pronounced as the bias increases.
(Right) The non-linear and bias contributions to the cross-correlation
coefficient, $\rho=P_{hm}/\sqrt{P_{hh}P_{mm}}$, as a function of $k$
(see Eq.~\ref{eqn:rho2}) assuming no shot noise.
The difference of $\rho^2$ from unity grows to high $k$ due to non-linear
structure formation (the ``1'' term) and the complexities of bias (the other
terms) as discussed in \S\ref{sec:Pmm}.}
\label{fig:bofk}
\end{figure}

Fig.~\ref{fig:bofk} shows the scale dependence of the bias estimated from
the cross- and auto-spectra.  As the redshift increases and halos of a fixed
mass become more biased the scale dependence of the bias becomes more
pronounced and the scale dependence differs more markedly between the auto-
and cross-spectra.
This means that a model based purely upon the dark matter power spectrum
becomes increasingly less accurate, even though that spectrum itself is
better approximated by linear theory.

\section{Measuring $P_{mm}(k,z)$}
\label{sec:Pmm}

A proper accounting of the growth of large scale structure through time is
one of the main goals of observational cosmology.  A key quantity in this
program is the matter power spectrum as a function of redshift.
Here we discuss how the cross-correlation can be combined with
the convergence or tracer auto-correlation to measure $P_{mm}(k,z)$.
To illustrate this measurement we pretend that the amplitude of the linear
theory spectrum ($\sigma_8$) was unknown, holding its shape fixed for
simplicity\footnote{It is easy to relax this assumption, but we want to vary
as few parameters as possible.  Note that some recent measurements of
$f\sigma_8$ with redshift-space distortions have also held the shape of the
power spectrum fixed \cite{Rei12}.}, and then attempt to recover its
value from the mock data.

First we model the auto and cross angular power spectra of a particular
sample in an individual redshift slice.  Following \S\ref{sec:zeff} we
take the spectra to be redshift independent over the width of the slice,
and assume that $dN/dz$ is perfectly known.
In such a situation we may schematically think of measuring the matter
power spectrum as:
\begin{equation}
  P_{mm}(k)\sim\frac{\left[C^{mg}_{\ell=k\chi}\right]^2}{C^{gg}_{\ell=k\chi}}
\end{equation}
Operationally we perform a joint fit to the combined data set, including
correlations and possibly some parameters to account for systematic errors.
With only the auto-spectrum there is a strong degeneracy between $\sigma_8$
and the bias parameters (particularly $b_1$).  However the matter-halo
cross-spectrum has a different dependence on these parameters and this
allows us to break the degeneracy and measure $\sigma_8$.

We compare the ability of two models to fit $C_\ell^{\kappa g}$ and
$C_\ell^{gg}$ with errors appropriate to three different experimental
configurations.  Our fiducial setup is
(1) that of proposed future experiments with depth equivalent to the LSST
gold sample, which corresponds to the limiting magnitude $i_{\rm lim} = 25.3$,
combined with a CMB experiment having $1\,\mu$K-arcmin noise and a beam of
$1.5'$.
To see how noise in imaging and CMB surveys impact our fits, we also do the
analysis for experimental configurations corresponding to 
(2) higher shot noise, which is modeled by using a limiting magnitude
$i_{\rm lim}=24.3$ while keeping CMB noise fixed, and another setup
corresponding to
(3) a CMB experiment having $5\,\mu$K-arcmin noise and a beam of $3'$
with $i_{\rm lim}=25.3$.
We always assume the overlap of the CMB and imaging surveys is
$f_{\rm sky}=0.5$.  Our errors scale as $f_{\rm sky}^{-1/2}$.

We have generated mock data, $C_\ell^{\kappa g}$ and $C_\ell^{gg}$,
assuming $P_{mh}(k)$ and $P_{hh}(k)$ from our N-body simulations at the
central redshift of our slice, with $dN/dz$ appropriate to LSST survey 
and correspoding $i_{\rm lim}$. We work in redshift slices of 
$\Delta z=0.5$ around the central redshift 
(e.g.~$1.75<z<2.25$ for $z=2$) and fit these mock data using two models.

Our fiducial model is the perturbation theory described in \S\ref{sec:lpt},
allowing $\sigma_8$, $b_1$, $b_2$, $\alpha_\times$ and $\alpha_a$ to vary.
As a comparison, and because it has been so widely used in the literature,
we use a model based on the {\sc HaloFit} fitting function for $P_{mm}(k)$.
As expected from Fig.~\ref{fig:nbody}, assuming a scale-independent bias is
insufficient to analyze any of the experimental configurations we consider --
the results are biased by many standard deviations.
To give some flexibility we allow the bias to be scale-dependent.
One choice, motivated by peaks theory, is to use $b(k)=b_{10}^E+b_{11}^Ek^2$
\cite{Des16}
where we have superscripted the $b_{ij}$ with an $E$ to indicate their Eulerian
nature and to distinguish them from the $b_i$ in Eq.~(\ref{eqn:pcross}).
We found that this choice alone does not provide a good fit to our N-body data,
as expected from Fig.~\ref{fig:bofk}.  Motivated by Fig.~\ref{fig:bofk}, but
as a purely phenomenological choice, we add a term linear in $k$ to our bias
model.  Then
\begin{eqnarray}
  P_{mh}(k) &=&\left[b_{10}^E+b_{1\frac{1}{2}}^Ek+b_{11}^Ek^2\right]  P_{HF}(k)
  \nonumber \\
  P_{hh}(k) &=&\left[b_{10}^E+b_{1\frac{1}{2}}^Ek+b_{11}^Ek^2\right]^2P_{HF}(k)
\label{eqn:halofit_peaks}
\end{eqnarray}
with $P_{HF}$ the {\sc HaloFit} fitting function for the matter auto-spectrum
and free parameters $\sigma_8$, $b_{10}^E$, $b_{1\frac{1}{2}}^E$ and $b_{11}^E$.

\begin{figure}
\begin{center}
\resizebox{\columnwidth}{!}{\includegraphics{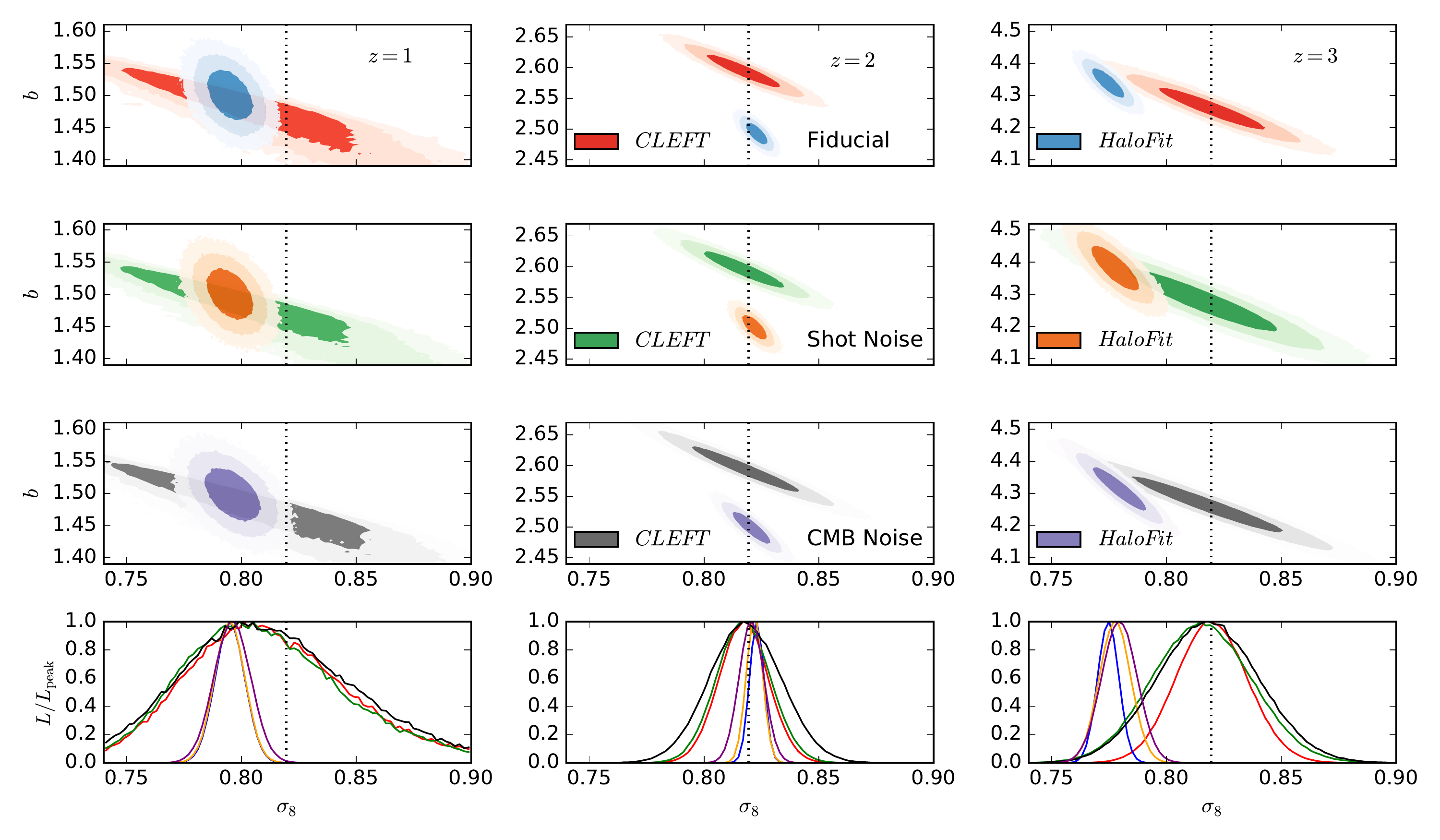}}
\end{center}
\vspace{-5mm}
\caption{Performance of the {\sc CLEFT} perturbation theory model and the
phenomological {\sc HaloFit} model at 3 different redshifts, $z=1$, $2$,
and $3$.  The top row show marginalized parameter distributions of $b$
and $\sigma_8$ for fits to $C_\ell^{\kappa g}$ and $C_\ell^{gg}$ for proposed
future experiment, while the second and third rows show the same distributions
but for increased shot noise and CMB experimental noise respectively (see text).
The fits are restricted to $\ell<2000$.
The definition of $b$ is different for the two models, so the values should
not be compared directly.
The last row shows the normalized posterior for $\sigma_8$ normalized to 
be 1 at the peak, with the vertical, black, dotted line marking the
 ``true'' value ($\sigma_8=0.82$) used in the simulations.}
\label{fig:summary}
\end{figure}

To evaluate the posteriors we run Monte Carlo Markov Chains (MCMC)
using the {\bf emcee}\footnote{https://github.com/dfm/emcee} 
package \cite{emcee} for both models at $z=1$, $2$ and $3$, for
all three experimental configurations. Unless specified otherwise, we
will quote the bias in the $50^{\rm th}$ percentile values (i.e.~median)
of these fits and $1\,\sigma$ errors based on the $16^{\rm th}$ and
$84^{\rm th}$ percentile values, but we have verified that using other
statistics such as mean or standard deviations does not change the numbers. 
We restrict the fits to $\ell_{\rm max}=2000$, even though the experiments
have useful measures of $C_\ell^{\kappa g}$ and $C_\ell^{gg}$ beyond this
value.  This is based on the discussion in \S\ref{sec:lpt} and also because
we found that going to higher $\ell$ does not improve the fits significantly 
(see below).

Fig.~\ref{fig:summary} compares the marginalized likelihoods in the
$\sigma_8-b$ plane (for the perturbation theory model we define $b=1+b_1$ 
while for the phenomenological model $b=b_{10}^E$).
At $z=2$ and $3$ the {\sc CLEFT} model returns unbiased constraints on
$\sigma_8$ for $\ell_{\rm max}=2000$.  In fact, we find that we can extend
the fits to $\ell_{\rm max}=3000$ without biasing our recovered $\sigma_8$
by $1\,\sigma$. Including higher $\ell$ reduces the $1\,\sigma$ errors from
$1.25\%$ to $1\%$ at $z=2$, but it also increases the bias from $0.3\%$ to
$0.5\%$.  At $z=3$, the errors are $1.8\%$ and $1.6\%$ with a bias of $0.05\%$
and $0.1\%$ respectively for the two $\ell_{\rm max}$.
For $z=1$ the {\sc CLEFT} model is biased whenever $\ell_{\rm max}>500$,
which is not unexpected given the discussion in \S\ref{sec:nbody}.  We expect
this bias would increase further if we pushed below $z=1$.
We shall discuss improvements to the model which could extend the reach to
lower redshift later.

The {\sc HaloFit} model provides much tigher constraints on $\sigma_8$ than
{\sc CLEFT} ($1\,\sigma$ errors of $0.34\%$ compared to $1.25\%$ at $z=2$),
however the estimates are biased by many $\sigma$ when fit to the same
$\ell_{\rm max}=2000$ as {\sc CLEFT} ($0.63\%$, or $2\,\sigma$ compared to
$0.33\%$, or $(1/3)\sigma$, at $z=2$).
This is initially surprising, given that the claimed $k$-range of validity of
{\sc HaloFit} is larger than for perturbation theory, but reinforces the
necessity of a sophisticated bias model and the high level of precision
demanded of fitting functions if they are to be used to interpret future data.
We find that at $z=2$ it is possible to get an unbiased estimate\footnote{The
more normal bias form, $b=b_{10}^E+b_{11}^E\,k^2$, also returns an unbiased
value of $\sigma_8$ as well at $z=2$ if we restrict the fit to $\ell<1500$
but is $5\,\sigma$ off if we use $\ell_{\rm max}=2000$.} of $\sigma_8$ by
reducing $\ell_{\rm max}$ to $1500$, however even this is not sufficient at
$z=1$ and $3$.  In fact, at $z=1$ {\sc HaloFit} breaks down at the same scale
as {\sc CLEFT} ($\ell=500$): the central value is as biased as for {\sc CLEFT}
but since the error bar is significantly smaller the central value is
$3\,\sigma$ away from the truth.

We also find that at $z=2$, where the galaxy auto-clustering is well above
the shot noise, going one magnitude shallower does not significantly increase
the uncertainty on $\sigma_8$ (from $1.25\%$ to $1.36\%$).
Our fits are more sensitive to CMB noise.  Increasing the noise to
$5\,\mu$K-arcmin increases our errors to $1.75\%$. 
However at $z=3$, where the survey becomes shot noise dominated at
$\ell>1800$, we are equally sensitive to CMB noise and increased shot noise,
with $1\,\sigma=2.5\%$ for both of them compared to $1.8\%$ for the
fiducial survey.

The performance of {\sc HaloFit} times a polynomial bias function clearly
highlights the necessity of a more sophisticated modeling approach in order
to make use of the massive amounts of cosmological information which will be
provided by future CMB and imaging surveys.
Even with the additional linear term introduced to better model bias,
{\sc HaloFit} is not able to fit scales where the data still have significant
constraining power.
Further, we note that for halos of $M_h\simeq 10^{12}\,h^{-1}M_\odot$, $z=2$
seems to be a sweet spot where the matter distribution is not highly non-linear
while at the same time the observed tracers are not very highly biased.
We expect such a sweet spot to exist for any halo mass, since halos are more
biased to higher redshifts while the clustering is more non-linear to lower
redshifts.  To push the sweet spot to higher redshift requires selecting lower
mass halos, which generally host lower luminosity galaxies.
Given these factors it is not surprising that the performance of {\sc HaloFit}
deteriorates in either direction from $z=2$.
On the other hand, that the performance of {\sc CLEFT} remains more or
less unchanged on going from $z=2$ to $z=3$, suggests that the bias model
employed is already flexible enough, even though we have not used the
additional bias parameters $b_{s^2}$ and $b_{\nabla^2}$.
This is also suggested by the fact that the {\sc CLEFT} best fits for
$\sigma_8$ are obtained when both $C_\ell^{\kappa g}$ and $C_\ell^{gg}$
are fit to same $\ell_{\rm max}$.  The quadratic dependence on bias in
$C_\ell^{gg}$ compared to the linear dependence in $C_\ell^{\kappa g}$ does
not lead to break down at lower $\ell$.
By working to the same $\ell_{\rm max}$ for both statistics we are able to
better break the parameter degeneracies.

\begin{figure}
\begin{center}
\resizebox{\columnwidth}{!}{\includegraphics{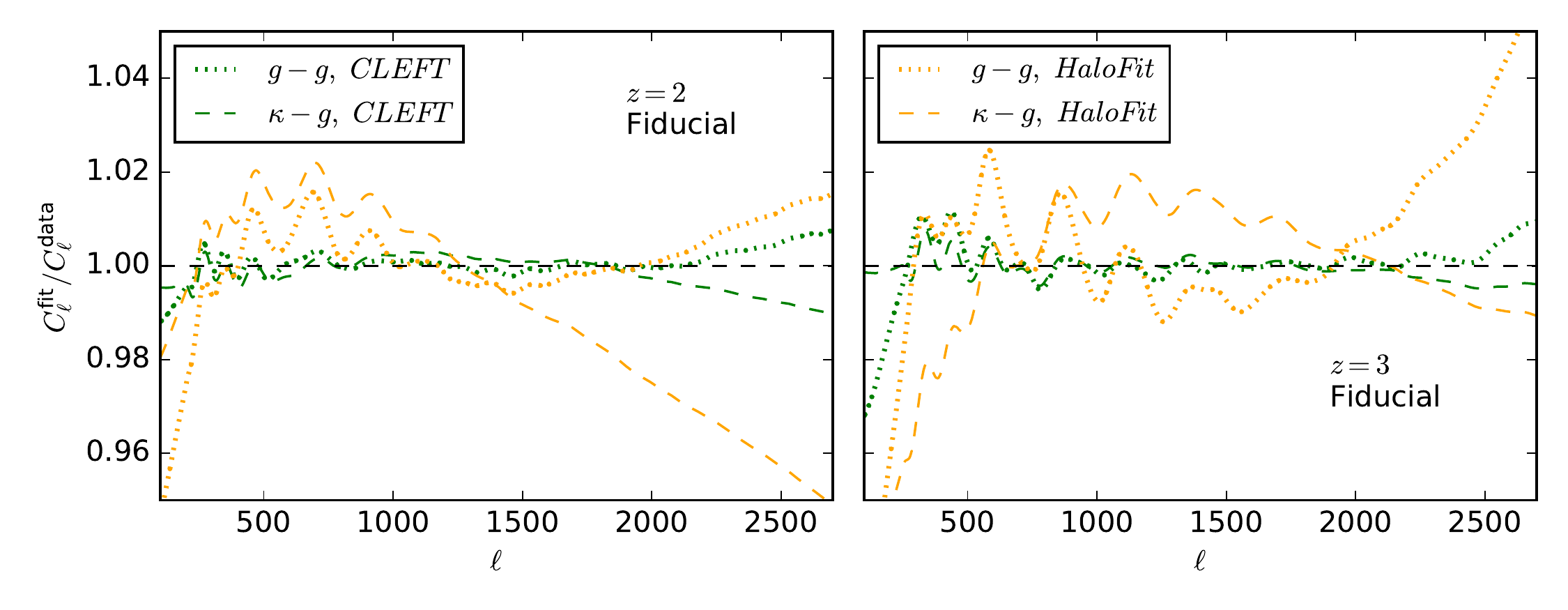}}
\end{center}
\vspace{-8mm}
\caption{Comparison of the fit to $C_\ell^{\kappa g}$ and $C_\ell^{gg}$ and
the data for the best fits of the {\sc CLEFT} model and the phenomological
{\sc HaloFit} model at $z=2$ (left) and $3$ (right) for our fiducial
experiment (see text).  The fits use $\ell_{\rm max}=2000$.}
\label{fig:summarycl}
\end{figure}

In Fig.~\ref{fig:summarycl} we also compare the data and the fits at the 
level of power spectra.  We show the best fits at $z=2$ and $z=3$ for the
fiducial experiment.
Despite fitting only up to $\ell_{\rm max}=2000$, the best fitting {\sc CLEFT}
power spectra are well within $1\%$ of the data on all scales of interest
($200<\ell<2700$).  This is below the statistical error in the data.
Such good agreement may be partly coincidence and reflect some over-fitting,
but reinforces how robust the results we obtain are to the exact choice of
fitting range etc.
By contrast the {\sc HaloFit} fits start to diverge beyond $\ell\simeq 1800$
and have $\sim 2\%$ excursions on intermediate scales.
The fits are qualitatively similar for other cases, which we omit for brevity.
Another view of the impact of the differences highlighted in
Fig.~\ref{fig:summarycl} is that the $\chi^2$ of the best fitting
{\sc CLEFT} models is 40 (60) lower than that of the best fitting {\sc HaloFit}
models at $z=2$ (3) despite having only one additional degree of freedom.

The {\sc CLEFT} model has several free parameters and it is straightforward
to see the cost that is paid in terms of error budget due to marginalizing
over extra parameters.
In Fig.~\ref{fig:fisher}, we show the error in $\sigma_8$ as the function of
$\ell_{\rm max}$ to which we fit the model.  We always marginalize over the
EFT parameters, but investigate the impact of tight priors on the bias
parameters.  The model used above corresponds to the green curve, marginalizing
over $b_1$, $b_2$, $\alpha_a$ and $\alpha_\times$.
We note that including one extra bias parameter ($b_2$) over linear bias
($b_1$) does not increase the error more than $0.5\%$.
As the above discussion makes clear, however, a proper bias model does
drastically reduce the bias in the fits (not shown in this Fisher calculation).
The situation changes as we marginalize over additional bias parameters,
e.g.~$b_{s^2}$.  Due to the degeneracies introduced, the fit to any given
$\ell$ becomes less constraining.

The above is the most natural combination of data for photometric surveys
at high $z$.  For completeness we remark upon two other possibilities.
(1) If redshifts are available for the tracer sample then one can
fit the multipoles of the redshift-space power spectrum in order to obtain
better constraints on the parameters and an independent constraint on the
amplitude.  The formalism of \S\ref{sec:lpt} allows such a fit within the
same parameter-set as the current study.  One advantage of using the 3D
clustering is that there are more modes\footnote{For the same volume there
are $(\ell_{\rm max}/\pi)(\Delta\chi/\chi)=(k_{\rm max}\Delta\chi/\pi)$
more modes in a 3D survey than a 2D survey that probes to the same
$\ell_{\rm max}$.  Note that for $\Delta z=0.5$ the ratio $\Delta\chi/\chi$
ranges from 37\% at $z=1$ to 8\% at $z=3$.} so we can work at larger scales
with the same statistical constraining power.
Another advantage is that the anisotropy of the clustering gives another
measure of $\sigma_8$.
A disadvantage is the need to model effects such as fingers-of-god.
We leave such an investigation to future work.

(2) Another route to measuring the power spectrum amplitude, though without
the redshift specificity, is through $C_\ell^{\kappa\kappa}$.
While the auto-spectrum of the tracers is likely to have higher
signal-to-noise ratio than the auto-spectrum of $\kappa$ it may be that
systematics in the tracer spectrum or complications of the bias model
favor using $C_\ell^{\kappa\kappa}$.
In this case the perturbation theory of \S\ref{sec:lpt} is not directly
applicable, since the integral for $C_\ell^{\kappa\kappa}$ probes low
redshifts and high $k$ values.  However, if the low $z$ contribution to
$C_\ell^{\kappa\kappa}$ can be ``cleaned'' by using a tracer of the density
field at low redshift (e.g.~LSST galaxies) then the power spectrum of the
cleaned map may be amenable to computation using our formalism.
Such a cleaned map may also have smaller contributions from intrinsic
bispectrum terms due to non-linear structure formation \cite{Bohm16}.

If we clean the $\kappa$ map using a biased tracer, the power spectrum of
the residual field can be written
\begin{equation}
  C_\ell^{\rm clean} = \sum_a C_{\ell,a}^{\kappa\kappa}\left(1-\rho_a^2\right)
\end{equation}
where $C_{\ell,a}^{\kappa\kappa}$ is the contribution to
$C_\ell^{\kappa\kappa}$ from redshift slice $a$ and
$\rho_a^2=(C_{\ell,a}^{\kappa g})^2/C_{\ell,a}^{\kappa\kappa}C_{\ell,a}^{gg}$
is the cross-correlation coefficient (squared) for slice $a$.
The $C_{\ell,a}^{gg}$ appearing in $\rho^2$ is to be interpreted as a ``total''
spectrum, including shot noise, such that having no galaxies in the slice
sends $\rho\to 0$.
We can estimate $1-\rho^2$ using our perturbative model.  If we treat the EFT
and bias terms perturbatively, in the spirit of Ref.~\cite{VlaCasWhi16}, then
$1-\rho^2$ only has contributions from 1-loop terms
\begin{equation}
  \rho^2\approx\frac{(P^{mh})^2}{P^{mm}P^{hh}}
        = 1 - \frac{b_1^2}{(1+b_1)^2}\sum_A \rho_A^2\frac{P_A}{P_Z}
            - \frac{{\rm const}}{(1+b_1)^2 P_Z}
\label{eqn:rho2}
\end{equation}
where $\rho_A^2$ are coefficients and $P_A$ are 1-loop power spectra,
both given in Table~\ref{tab:rho2}.
The last term, containing the ``const'', is the leading stochastic contribution 
obtained by expanding the terms in $P^{hh}$.  Alternatively, this leading 
stochastic (constant) contribution could be treated non-perturbatively, 
keeping it in the denominator of the $1/P^{hh}$ expansion, which would in  
turn affect all the rest of the terms in the sum above.
Note that the degree of decorrelation is dependent upon the non-linear nature
of the bias model, small-scale physics and the shot noise, as expected.
The quantities $P_A$ are all 1-loop in our perturbation expansion, and
$P_A/P_Z\to 0$ as $k\to 0$ indicating that $\rho^2\to 1$ on large scales if
the shot-noise is sufficiently small.  The fields begin to decorrelate when
the 1-loop corrections become important, and the degree of decorrelation is
larger when the objects are more biased (as expected).
Again, we leave investigation of this possibility to future work.

\begin{table}
\begin{center}
\begin{tabular}{ccc|ccc}
$A$          & $\rho_A^2$ & $P_A$ &
$A$          & $\rho_A^2$ & $P_A$ \\ \hline
$1$          & 1
             & $P_{1-{\rm loop}}$   &
$b_{s^2}$    & $b_{s^2}/b_1$
             & $2P_{b_1b_{s^2}}-P_{b_{s^2}}$ \\
$b_1$        & -1
             & $P_{b_1}-2P_Z$       &
$b_2b_{s^2}$ & $b_2b_{s^2}/b_1^2$
             & $P_{b_2b_{s^2}}$     \\
$b_1^2$      & 1
             & $P_{b_1^2}-P_Z$      &
$b_{s^2}^2$  & $(b_{s^2}/b_1)^2$
             & $P_{(b_{s^2})^2}$    \\
$b_2$        & $b_2/b_1$        
             & $P_{b_1b_2}-P_{b_2}$ &
$\alpha$     & $2\alpha_\times-\alpha_m-\alpha_a$ 
             & $k^2P_Z$             \\
$b_2^2$      & $(b_2/b_1)^2$        
             & $P_{b_2^2}$          &&        
\end{tabular}
\caption{Coefficients, $\rho_A^2$, and 1-loop power spectra, $P_A$, for
the correlation coefficient of Eq.~(\ref{eqn:rho2}).}
\end{center}
\label{tab:rho2}
\end{table}

\section{The redshift distribution}
\label{sec:dndz}

\begin{figure}
\begin{center}
\resizebox{\columnwidth}{!}{\includegraphics{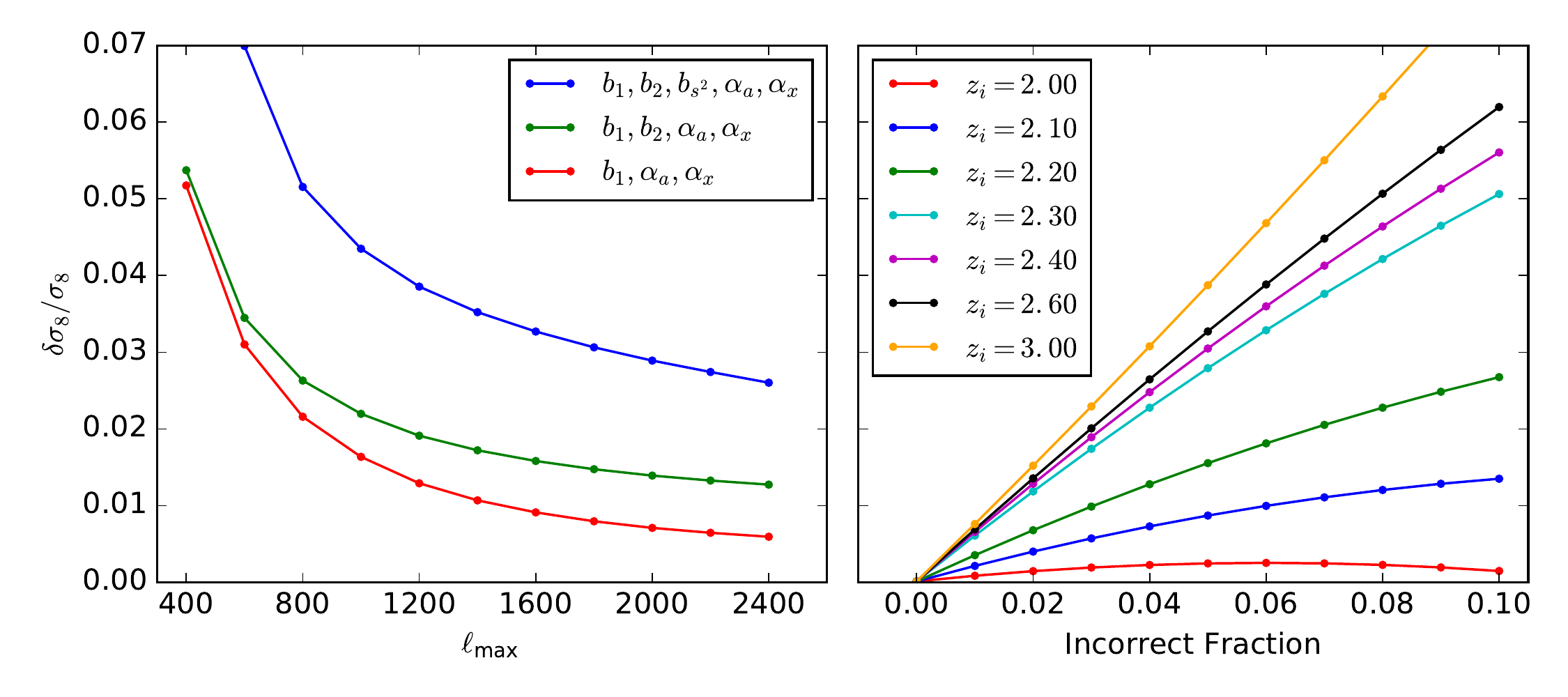}}
\end{center}
\caption{Fisher analysis to predict error on $\sigma_8$. The left panel shows 
hows the error incurred in $\sigma_8$ due to marginalizing over extra parameters
that are used in {\sc CLEFT} model as a function of $\ell_{\rm max}$ to which
we fit the data. 
Right panel shows the error in $\sigma_8$ due to incorrect $dN/dz$, modeled by
adding a Gaussian bump of $FWHM=0.1$ in $z$ centered at redshift $z_i$ to
the fiducial LSST $dN/dz$ in a slice of width $\Delta z= 0.5$ centered at
$z=2$. The $x$-axis shows the fraction of sources misidentifed in the LSST
sample.}
\label{fig:fisher}
\end{figure}

One of the advantages of a cross-correlation is that it isolates the
contribution to the lensing signal arising from a small redshift interval,
enabling a study of redshift evolution.
Such an analysis, however, relies on being able to choose sources which
have some known or desired redshift distribution.
In this section we look at how accurately we need to determine $dN/dz$ in
order to not be limited by this uncertainty.

A change in $dN/dz$ has two effects: it changes the mix of redshifts which
contribute to $C_\ell$ and it changes the mix of scales ($k$ values) which
contribute to a given $\ell$.  The change to $C_\ell^{\kappa g}$ is linear
in $dN/dz$ while the change in $C_\ell^{gg}$ is quadratic.
We will assess the impact of these changes using a linear approximation,
where an error in $dN/dz$ is assumed to be small.  In the small-error limit
a bias in the data, $\delta d_n$, leads to biases in the parameters,
$p_\alpha$, of:
\begin{equation}
  \delta p_\alpha=F^{-1}_{\alpha\beta}
  \frac{\partial\mu_m}{\partial p_\beta} C^{-1}_{mn} \delta d_n
\label{eqn:fisher}
\end{equation}
where $C$ is covariance matrix of the data, $\mu$ is the model prediction 
and $F$ is the corresponding Fisher matrix.
In the spirit of the last section, we shall focus on the bias introduced in
$\sigma_8$ assuming
\begin{equation}
  \frac{dN}{dz} = (1-f)\left(\frac{dN}{dz}\right)_1
                    +f \left(\frac{dN}{dz}\right)_2
\end{equation}
where $(dN/dz)_2$ is offset from $(dN/dz)_1$ by a varying amount and $f$ is
the fraction of sources misidentified in the survey.
The largest derivatives, $d\mu_m/dp_\alpha$, are for $p=\sigma_8$ and $b_1$
so we shall specialize to the $2\times 2$ Fisher matrix.

Our toy model for $dN/dz$ is to take $(dN/dz)_1$ as the LSST-like distribution
cut to $1.75<z<2.25$ while introducing a Gaussian $(dN/dz)_2$ of FWHM $=0.1$
centered at $z_i$.  We use the N-body $P(k,z)$ of our $10^{12}\,h^{-1}M_\odot$
halo sample to calculate $\delta d_n$ and propagate the bias to $\sigma_8$
using Eq.~(\ref{eqn:fisher}).
Fig.~\ref{fig:fisher} shows this fractional error as a function of fraction
of misidentified sources (defined as the fraction of total galaxies in the
Gaussian), at different values of $z_i$. We expect these errors to 
asymptote with increasing $z_i$, since in the extreme case of 
$z_i \rightarrow \infty$, the error in $\delta d_n$ should be equal to
the error due to reduction in total number of sources.

We note in passing that our formalism could in principle be used to
constrain $dN/dz$ at the same time as fit for the model parameters.
Since the formalism naturally encompasses cross-correlations between
biased tracers in real and redshift space a particularly interesting
case would be to constrain $dN/dz$ through cross-correlations with a
spectroscopic survey (e.g.~Ref.~\cite{McQWhi13} and references therein)
while simulataneously fitting the $\kappa-$tracer statistics.
We leave investigation of this possibility for future work.

\section{Conclusions}
\label{sec:conclusions}

A new generation of deep imaging surveys and CMB experiments offers the
possibility of using cross-correlations to test General Relativity, probe
the galaxy-halo connection and measure the growth of large-scale structure.
However improvements in data require concurrent improvements in the
theoretical modeling in order to reap the promised science.
We have investigated the use of Lagrangian perturbation theory to model
cross-correlations between the lensing of the CMB and biased tracers
of large-scale structure at high $z$.

Ever lower map noise levels improve the fidelity of CMB lensing maps,
with the improvement becoming particularly significant once the noise
in the foreground-cleaned maps reaches $\sim 2\,\mu$K-arcmin and the
$EB$ spectra dominate.
With such improvements maps of the lensing convergence will go from being
noise dominated above $\ell\sim 10^2$ to noise dominated only above
$\ell\sim 10^3$, an increase of two orders of magnitude in the number of
high signal-to-noise modes and hence useable information.
On a similar timescale dramatic increases in the depth and fidelity of
optical imaging over large sky areas will come from a next generation of
surveys, allowing probes of higher redshift galaxies where the CMB lensing
kernel peaks.  The combination of these two advances enables multiple
science goals through cross-correlations.

We have argued above that the particular scales and redshifts which
contain much of the cosmological information in cross-correlations can be
modeled using cosmological perturbation theory.  This extends the highly
successful linear perturbation theory analysis of primary CMB anisotropies
which has proven so impactful.  It provides a first-principles approach
with a sophisticated treatment of bias for the halos and galaxies which are
directly observed by the imaging surveys.

In fact, a flexible and sophisticated bias model is critically important
in modeling CMB lensing-galaxy cross-correlations.  We show that the
commonly used scale-independent bias times matter power spectrum approach
will be completely inadequate to analyze upcoming surveys, and that simply
extending the bias to a polynomial in $k$ does not solve the issue.
Rather a proper modeling of the non-linear effects of bias is essential.
In fact, in many ways a proper accounting for the complexities of bias is
more important than the effects of non-linear structure growth on the matter
power spectrum at high $z$ where the CMB lensing kernel peaks.
Since the non-linear scale shifts to smaller scales at high $z$, while the
Lagrangian radius of a fixed mass halo remains constant, the complexities
of bias will only become more relevant at higher $z$.

Comparing the clustering of halos in a series of N-body simulations to our
perturbative model, we found that the auto- and cross-clustering of halos
above $z=2$ could be well described up to $\ell=2000$ using only two
(Lagrangian) bias parameters.  While the formalism has been extended to
include higher order terms, they were not necessary for the tracers and
scales we investigated.

As an example of the science enabled by cross-correlations, we reconstructed
the amplitude of the matter power spectrum ($\sigma_8$) from the combination
of $C_\ell^{\kappa g}$ and $C_\ell^{gg}$ for some hypothetical experiments.
We found that constraints on $\sigma_8$ improved slowly beyond $\ell=2000$,
and that our fits became biased if we fit the N-body data to higher $\ell$.
Unless the modeling can be improved, or if the scale dependence of the bias
is partially degenerate with changes due to $\sigma_8$ above $\ell=2000$,
gains in experimental sensitivity at high $\ell$ will not advance this science
goal.  For our $10^{12}\,h^{-1}M_\odot$ halo sample and slices of width
$\Delta z=0.5$ the optical survey is shot-noise limited at $\ell=2000$ for
$160$, $390$ and $680$ galaxies per deg${}^2$ at $z=1$, 2 and 3.
The CMB lensing becomes noise dominated beyond $\ell=10^3$ at sensitivities of
$\sim 5\,\mu$K-arcmin for a wide range of beam sizes.
While it is relatively forgiving of map depth and angular resolution, like
most cross-correlation science the uncertainty scales as $f_{\rm sky}^{-1/2}$,
preferring large sky coverage, overlapping other surveys.
We found that cross-correlations of the sort enabled by LSST and CMB-S4 would
enable percent level measurements of $\sigma_8$ in multiple redshift bins.
Deeper imaging would allow us to extend these measurements all the way to
$z\simeq 6$.

We have shown that a scale-independent, or ``linear'', bias does not provide
a good model for $k>0.1-0.2\,h\,{\rm Mpc}^{-1}$ at any redshift we have
studied (see Fig.~\ref{fig:nbody}).
This is at odds with the assumptions of Ref.~\cite{Sch17} in their forecasts.
Those authors assume linear bias works to $k\simeq 0.6\,h\,{\rm Mpc}^{-1}$ at
all redshifts ($2\pi\chi/\ell>10\,h^{-1}$Mpc).  While the conclusions of
Ref.~\cite{Sch17} on shear calibration are very likely unchanged (they can
achieve the LSST shear calibration requirements with only cross- and
auto-spectra of shears in any case) it would be interesting to revisit these
forecasts with a more flexible bias model.  Such an investigation, and an
exploration of the degeneracies introduced, is beyond the scope of this paper 
although we give some additional comments below.

The contributions of the different biasing terms in Eq.~(\ref{eqn:pcross})
are formally independent, however if we keep all of the terms there is a
danger of over-fitting.  For a fixed precision, e.g.~1\%, the different
contributions can exhibit approximate degeneracies so that a subset of the
terms can mimic the effects of the rest.
In principle adding higher order statistics, like the bispectrum, can help
break approximate degeneracies.  In absence of such additional information,
an alternative is to reduce a number of independent terms.
This is the approach we have adopted in our analysis: keeping the $b_1$ and
$b_2$ bias parameters.  We note that the values of these parameters should
now be understood in the `effective' sense, since they also partially take
the role of $b_{s^2}$ and $b_{\nabla^2}$ terms, which could additionally
change the numerical values of these parameters from the peak-background
split estimates given by Eq.~(\ref{eq:ps_bias}).

In addition to the biasing parameters we have considered so far,
effects related to baryonic physics leaking in from the small scales
can affect the galaxy clustering even on fairly large scales.
Additionally, we can also consider relative-density and relative-velocity
perturbations that can also potentially appear on large scales. 
These baryonic effects can be added to the biasing description, considering
them as an additional species adding to the full set of symmetry allowed
terms for the galaxy overdensity \cite{Ang15,Sch16}, starting from
additional relative-density $\delta_{\rm bc}$ and relative-velocity
$\theta_{\rm bc}$ perturbation. 
In addition to these effects we have also higher order contributions
starting from the relative velocity effects \cite{Yoo13,Ang15,Bla16,Sch16},
though these terms have also been recently studied and constrained to be a
rather small effect \cite{Sle16,Beu16} relative to the rest of the terms.
We also note that similar effects described by the general formalism presented
in \cite{Ang15,Sch16} could also be adapted to describe effects of cosmic
neutrinos or fluctuating dark energy models \cite{Fas16} on mildly-nonlinear
scales.
We note that the {\sc CLEFT} formalism used in this paper can be readily
extended to include these additional biasing terms. 

Finally we remark on some directions for future development.
While we have focused here on a single population of tracers, there are
significant gains which can be had by using multi-tracer techniques
\cite{McDSel09,WhiSonPer09,SchSel17}.
The formalism described above can be straightforwardly extended to the
multi-tracer case, including the decorrelations which occur at high $k$.
The inclusion of low mass neutrinos into the formalism is straightforward,
and there are extensions for models with modifications to General Relativity
\cite{BosKoy16,AviCer17}.  In the near future the demands on the theory are
significantly relaxed, and simpler approximations can perform adequately.
We discuss one such approach in Appendix \ref{app:hz}.
In the other direction, we have focused throughout on 1-loop perturbative
predictions with a Lagrangian bias model which is $2^{\rm nd}$ order in the
linear density field.
We saw that at high redshift the uncertainties due to the bias model dominated
over the non-linearities in the matter clustering.  However this situation
changes as we move to lower redshift and the clustering becomes more non-linear.
There is no reason, in principle, why one cannot continue the expansion to
higher order in order to deal with this.
Ref.~\cite{BalMerZal15} presents the 2-loop EFT calculation (in Eulerian PT)
for the matter power spectrum.  At this order there are 6 EFT counter terms
which need to be fixed.  These terms are highly degenerate, and
Ref.~\cite{BalMerZal15} discuss some ways to reduce this number.
It is an open question whether one needs to work at higher order in the bias
expansion.  If so, this would add additional parameters.
However at lower redshift we expect to be using galaxies of lower bias and
we have the ability to adjust our samples so as to minimize the
scale-dependence of their bias so it is likely that we can stay with the same
order as used above.
If we keep all of the EFT terms then even at the same order in the bias
expansion we would be fitting 9 parameters to the data, rather than our
current 4 (plus $\sigma_8$).
It remains to be seen whether the increase in information afforded by going to
smaller scales using higher order perturbation theory overcomes the need to
introduce more free parameters in the fit.  On the other hand, one can use
priors from N-body simulations or fitting functions to partially constrain
some of these additional parameters, which would reduce the impact of their
degeneracies.  We leave this possibility for future work.

We will make our code publicly available\footnote{\tt
https://github.com/martinjameswhite/CLEFT\_GSM}.

\acknowledgments
We would like to thank Arjun Dey, Simone Ferraro, Emmanuel Schaan,
David Schlegel, Marcel Schmittfull, Uros Seljak and Blake Sherwin
for useful discussions during the preparation of this manuscript.

Z.V.~is supported in part by the U.S. Department of Energy contract to
SLAC no.~DE-AC02-76SF00515.
M.W.~is supported by the U.S. Department of Energy.

This research has made use of NASA's Astrophysics Data System.
The analysis in this paper made use of the computing resources of the
National Energy Research Scientific Computing Center.

\appendix

\section{Noise model for forecasts}
\label{app:noise}

We follow standard practice to estimate the precision with which
measurements of the cross-power spectrum can be measured.  Assuming
the fields are Gaussian, and specializing to the case of galaxy-lensing
cross-correlations, the variance on the cross-power is
\begin{equation}
  {\rm Var}\left[C_\ell^{\kappa g}\right] =
  \frac{1}{(2\ell+1)f_{\rm sky}} \left\{
  \left(C_\ell^{\kappa\kappa}+N_\ell^{\kappa\kappa}\right)
  \left(C_\ell^{gg}+N_\ell^{gg}\right) +
  \left(C_\ell^{\kappa g}\right)^2\right\}
\end{equation}
where $f_{\rm sky}$ is the sky fraction, $C_\ell^{ii}$ represent the signal
and $N_\ell^{ii}$ the noise in the auto-spectra.
If $C_\ell^{\kappa g}=r\sqrt{C_\ell^{\kappa\kappa}C_\ell^{gg}}$ then the
sample variance limit becomes
\begin{equation}
  \frac{\delta C_\ell^{\kappa g}}{C_\ell^{\kappa g}} \to
  \sqrt{\frac{2}{(2\ell+1)f_{\rm sky}}\ \frac{1+r^2}{2r^2}}
\end{equation}
and future observations will be sample variance limited to $\ell\simeq 10^3$.
For completeness, under the same assumptions the $gg$ variance is
\begin{equation}
  {\rm Var}\left[C_\ell^{gg}\right] =
  \frac{2}{(2\ell+1)f_{\rm sky}} \left(C_\ell^{gg}+N_\ell^{gg}\right)^2
\end{equation}
and the covariance between $C_\ell^{\kappa g}$ and $C_\ell^{gg}$ is
\begin{equation}
  {\rm Cov}\left[C_\ell^{\kappa g},C_\ell^{gg}\right] =
  \frac{2}{(2\ell+1)f_{\rm sky}} \left\{ \vphantom{\int_0^1}
  C_\ell^{\kappa g} \left(C_\ell^{gg}+N_\ell^{gg}\right) \right\}
\end{equation}

We model the noise in the galaxy autospectrum as simple shot noise, with
$N_\ell^{gg}=1/\bar{n}$ and $\bar{n}$ the angular number density of
tracers in the sample.  For the CMB lensing signal we make the usual
assumption that it is dominated by the fluctuations in the primary
CMB signal (and detector noise) and can be approximated by the signal-free
component \cite{HuOka02,Han10}.
Taking the flat-sky limit appropriate to high $\ell$
\begin{equation}
  N_L^{\kappa\kappa} = \left[\frac{\ell(\ell+1)}{2}\right]^2
  \left[\int\frac{d^2\ell}{(2\pi)^2} \sum_{(XY)}
  K^{XY}(\vec{\ell},\vec{L})\right]^{-1}
\end{equation}
where we have assumed full sky coverage, $(XY)$ denotes a sum over pairs
of $T$, $E$ and $B$ modes and $K^{XY}$ are kernels depending upon the angular
power spectra of the CMB \cite{HuOka02}.
In the above we truncate the integrals at $\ell_{\rm max}=3000$ for $TT$
$\ell_{\rm max}=5000$ for $EE$ and $EB$.
For future, low-noise experiments and at high $\ell$ we expect the measurement
to be dominated by the $EB$ cross-correlation,
\begin{equation}
  K^{EB}(\ell,L) =
  \frac{[(\vec{L}-\vec{\ell})\cdot\vec{L}C_{\ell-L}^B
  +\vec{\ell}\cdot\vec{L}C_\ell^E]^2}{C_\ell^{{\rm tot},E}
   C_{\ell-L}^{{\rm tot},B}}\sin^2(2\phi_\ell)
\end{equation}
with a smaller contribution from $TT$,
\begin{equation}
  K^{TT} = \frac{[(\vec{L}-\vec{\ell})\cdot\vec{L}C_{\ell-L} +
                   \vec{\ell}\cdot\vec{L} C_\ell]^2}
                {2C_\ell^{{\rm tot},T} C_{\ell-L}^{{\rm tot},T}}
\end{equation}
The $C_\ell^{\rm tot}$ include contributions from the lensed CMB and the
noise.  Recalling that the lensing-induced $C_\ell^B$ is approximately
constant at low $\ell$ and much less than $C_\ell^E$ we expect the $EB$ noise
on $\kappa$ to be scale-independent at low $\ell$
(since $C_\ell^B$ is negligible and instrumental noise is scale independent
on large scales).
Assuming fully polarized detectors with white noise
\begin{eqnarray}
  N_\ell^{T} &=& (\Delta_T/T_{CMB})^2\exp[\ell(\ell+1)\Theta_b^2/8\ln 2] \\
  N_\ell^{E} = N_\ell^{B} &=& (\Delta_P/T_{CMB})^2
  \exp[\ell(\ell+1)\Theta_b^2/8\ln 2]
\end{eqnarray}
with $\Theta_b$ the FWHM of the beam, $\Delta_P=\sqrt{2}\Delta_T$ and the
$\Delta_{T,P}$ given in K-radian (converted from $\mu\,$K-arcmin).

There are several assumptions in the above, which are probably adequate for
our forecasts of future experiments whose performance is anyway highly
uncertain.  We have neglected foreground subtraction, taking it into account
only in as far as we impose an $\ell_{\rm max}$ cut on the integrals.
One the other hand we have assumed the noise appropriate to the quadratic
estimator, though this is not optimal at very low noise and could be lowered
by using an iterative scheme (e.g.~Ref.~\cite{Smi12}).

In order to determine the shot-noise level for the optical survey we need
to know the number density of galaxies as a function of redshift.
We follow the LSST science book \cite{LSST} and assume
\begin{eqnarray}
  p(z) &=& \frac{1}{2\,z_0}\left(\frac{z}{z_0}\right)^2
        \exp\left[-\frac{z}{z_0}\right] \\
  z_0 &=& 0.0417\, i_{\rm lim} - 0.744
\end{eqnarray}
normalized to $N_g = 46\times 10^{0.3(i_{\rm lim}-25)}$ galaxies per
arcmin${}^2$.  We assume $i_{\rm lim}=25.3$ for the LSST gold sample.
This enables computation of $\bar{n}$ in any slice $(z;\Delta z)$.

\section{Simpler models}
\label{app:hz}

The requirements imposed by future imaging surveys and CMB experiments
upon the theoretical modeling are extreme.  However those surveys are
also in the future, and the requirements imposed by current generation
experiments are not as challening.  For this reason we examine here two
less accurate, but simpler, models for the auto- and cross-spectra of
biased tracers.

\begin{figure}
\resizebox{\columnwidth}{!}{\includegraphics{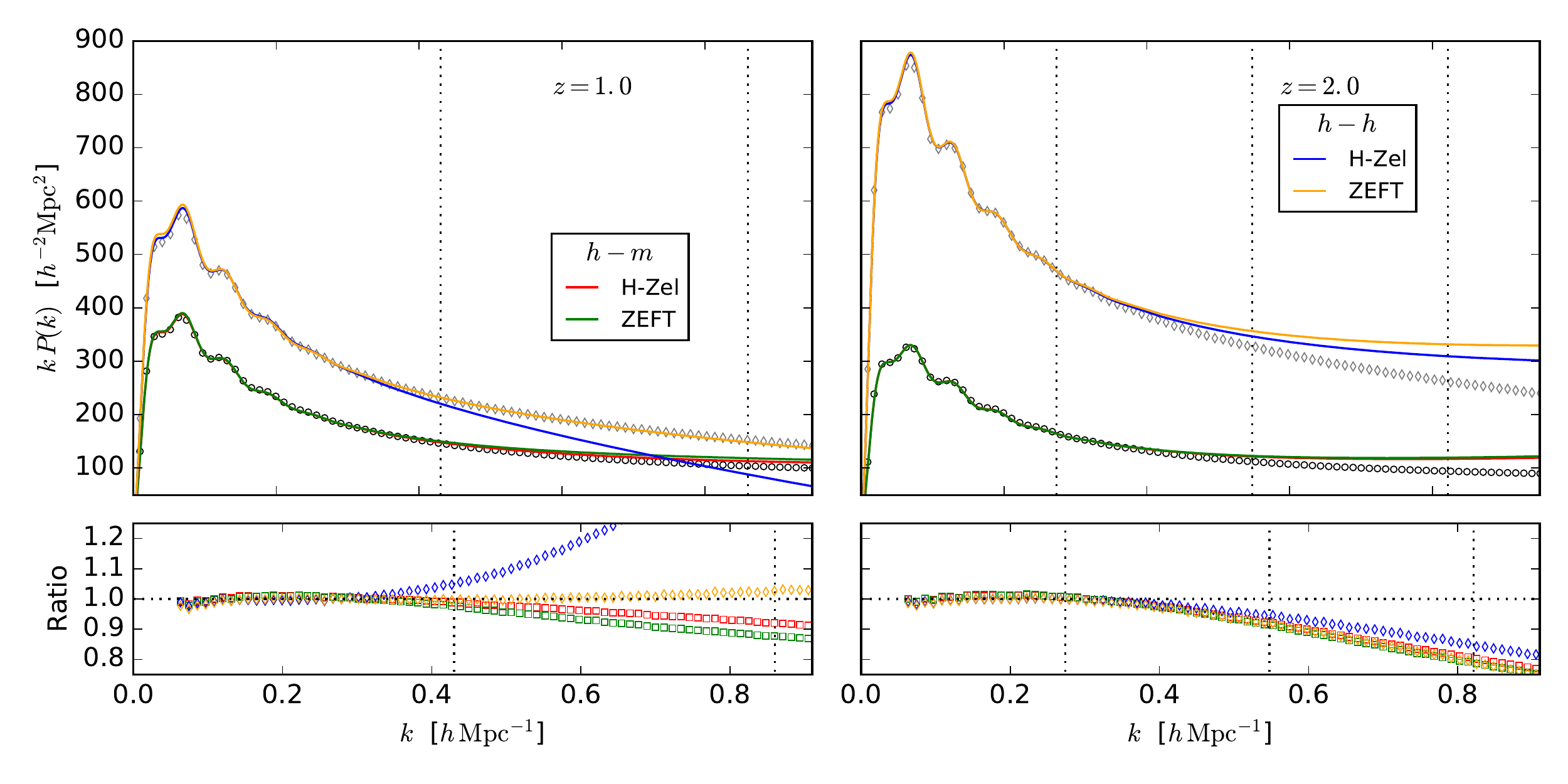}}
\caption{As for Fig.~\ref{fig:nbody}, but for the simpler models
Halo-Zeldovich (H-Zel) and ZEFT, defined using Eq.~(\ref{eqn:hz}).
The diamonds are for the halo auto-spectra while the circles
are for the halo-matter cross-spectra (we have
omitted the matter auto-correlation).  The lower panels now show the ratio
of the N-body to both the theories on an expanded $y$-axis scale for both.}
\label{fig:hz}
\end{figure}

The two simpler models are based upon the ``ZEFT'' model of
Ref.~\cite{VlaWhiAvi15} and the ``Halo-Zeldovich'' model of
Ref.~\cite{SelVla15, HandSelj17}, both modified to include
biased\footnote{Ref.~\cite{VlaCasWhi16} used a simple bias model purely
as a reference spectrum for plotting.  Here we use the more sophisticated
bias in order to be able to fit data.} tracers as in Ref.~\cite{Mat08b,Whi14}.
These models have the same number of free parameters as our CLEFT model and
the same flexible bias model, while requiring only 1D integrals of the
linear theory power spectrum so they can be evaluated very quickly.
While not as accurate as the CLEFT model, they provide an adequate fit to
our N-body simulations over the whole range of redshift (even at lower $z$)
which may be sufficient for the next few years.

The power spectra of the models is of the same form as Eq.~(\ref{eqn:Px}) but
the individual contributions, $f_x$, contain only tree-level terms.  For the
Halo-Zeldovich model there is a constant term added whose amplitude forms a
free parameter.  Explicitly
\begin{eqnarray}
  P_{HZ} &=& 4\pi\int q^2\,dq\ e^{-(1/2)k^2(X_L+Y_L)}\left\{\vphantom{\int}
  \right.  \nonumber \\
  & & \left[ 1
  + b_1^2\left(\xi_L-k^2U_L^2\right)
  - b_2  \left(k^2U_L^2\right)
  + \frac{b_2^2}{2}\xi_L^2 \right] j_0(kq) \nonumber \\
  &+& \sum_{n=1}^\infty \left[ 1
    -2b_1  \frac{q\,U_L}{Y_L}
    + b_1^2\left(\xi_L + \left[\frac{2n}{Y_L}-k^2\right]U_L^2\right)
    + b_2  \left(\frac{2n}{Y_L} -k^2\right)U_L^2 \right.
  \nonumber \\
  && \left. \left.
    - 2b_1b_2\frac{q\,U_L\,\xi_L}{Y_L}
    + \frac{b_2^2}{2}\xi_L^2
    \right] \left(\frac{k\,Y_L}{q}\right)^n j_n(kq) \right\}
    + 1-{\rm halo}
\label{eqn:hz}
\end{eqnarray}
where the integral over $q$ can be done efficiently using fast Fourier
transforms \cite{Tal78,Ham00} or other methods \cite{LucSto95,Oga05}
and we can take the 1-halo term as a constant.
For completeness
\begin{eqnarray}
  \xi_L(q)&=&\frac{1}{2\pi^2}\int_0^\infty dk\ P_L(k)
          \left[k^2\,j_0(kq)\right] \\
  X_L(q)  &=& \frac{1}{2\pi^2}\int_0^\infty dk\ P_L(k)
          \left[\frac{2}{3}-2\frac{j_1(kq)}{kq}\right] \\
  Y_L(q)  &=& \frac{1}{2\pi^2}\int_0^\infty dk\ P_L(k)
          \left[-2j_0(kq)+6\frac{j_1(kq)}{kq}\right] \\
  U_L(q)  &=& \frac{1}{2\pi^2}\int_0^\infty dk\ P_L(k)
          \left[-k\,j_1(kq)\right]
\end{eqnarray}
We have omitted the dependence upon $b_{s^2}$ and $b_{\nabla^2}$ as
unceccessary for this level of approximation.
As in the main text, the auto-correlation contains all of the terms while
the cross-correlation with the matter contains only terms linear in $b_1$
and $b_2$ (divided by 2).

The alternative is the ``ZEFT'' model.  In this model the 1-halo term is
replaced by $\alpha k^2P_{\rm Z}$, and thus has the same number of free
parameters.  Note that this substitution requires no additional calculation,
since $P_{\rm Z}$ is already computed as part of $P_{HZ}$.
Fig.~\ref{fig:hz} shows the performance of the models at $z=1$ and $2$
compared to our $10^{12}\,h^{-1}M_\odot$ halo sample.
We find the performance of the two models similar, with the ZEFT model
performing slightly better, especially at lower redshifts. Both the models
agree to within $1\%$ with the N-body results out to $k = 0.2\,h\,$Mpc$^{-1}$
and within $5\%$ to $k = 0.4\,h\,$Mpc$^{-1}$ at $z=1$ and $2$.
Neither model performs as well as CLEFT, even at high redshift, because
even at $z=3$ the 1-loop contributions to both, matter and the bias terms
are not negligible, and act to improve agreement with the N-body.

Of course it is possible to further improve the performance of these models
by introducing more free parameters or by combining the $\alpha k^2P_{\rm Z}$
and 1-halo terms.  We found this gave negligible improvement.  The 1-halo term
can be replaced by a power series in $k$, or a Pad\'{e} term.  We have
experimented with terms of the form $(k\Sigma)^2/[1+(k\Sigma)^2]$ but did not
find very dramatic improvements.  Further progress would obviously require
more degrees of freedom in the 1-halo term.

\bibliographystyle{JHEP}
\bibliography{ms}
\end{document}